\documentclass[showpacs,amsmath,amssymb,twocolumn,superscriptaddress,floatfix,prxquantum]{revtex4-2}
\usepackage[dvips]{graphicx}
\usepackage{enumerate}
\usepackage{epsfig}
\usepackage{xcolor}
\usepackage[T1]{fontenc}
\usepackage{fullpage}
\usepackage{amsthm,amsfonts,amscd,mathrsfs,xspace,framed}
\usepackage{color}
\usepackage{setspace}
\usepackage{url}
\usepackage{enumitem}
\usepackage{braket}
\usepackage{makecell}
\usepackage[dvipsnames]{xcolor}
\usepackage[colorlinks,citecolor=red,urlcolor=blue,bookmarks=false,hypertexnames=true]{hyperref}

\begin{document}

\title{Hardware-Efficient Universal Linear Transformations for Optical Modes in the Synthetic Time Dimension}

\author{Jasvith Raj Basani}
\thanks{These authors contributed equally.}
\affiliation{Department of Electrical and Computer Engineering, University of Maryland, College Park, Maryland 20742, USA}
\affiliation{Institute for Research in Electronics and Applied Physics, and Joint Quantum Institute, University of Maryland, College Park, Maryland 20742, USA}
\author{Chaohan Cui}
\thanks{These authors contributed equally.}
\email{chaohan@umd.edu}
\affiliation{Department of Electrical and Computer Engineering, University of Maryland, College Park, Maryland 20742, USA}
\affiliation{James C. Wyant College of Optical Sciences, University of Arizona, Tucson, Arizona 85721, USA}
\author{Jack Postlewaite}
\affiliation{Department of Electrical and Computer Engineering, University of Maryland, College Park, Maryland 20742, USA}
\author{Edo Waks}
\affiliation{Department of Electrical and Computer Engineering, University of Maryland, College Park, Maryland 20742, USA}
\affiliation{Institute for Research in Electronics and Applied Physics, and Joint Quantum Institute, University of Maryland, College Park, Maryland 20742, USA}
\author{Saikat Guha}
\affiliation{Department of Electrical and Computer Engineering, University of Maryland, College Park, Maryland 20742, United States}
\affiliation{James C. Wyant College of Optical Sciences, University of Arizona, Tucson, Arizona 85721, USA}


\begin{abstract}
Recent progress in photonic information processing has spurred strong demand in scalable and reconfigurable photonic circuitry. Conventional spatially-meshed multi-port interferometers require a number of components growing quadratically with the system size, posing a fundamental scaling challenge ahead. Here, we introduce a hardware-efficient synthetic time-domain photonic processor that achieves at least an exponential reduction in hardware component count for implementing arbitrary linear transformations. The processor’s dynamic connectivity allows systematic pruning, minimizing optical loss while preserving all-to-all connectivity. We benchmark our architecture on the task of boosted Bell state measurements -- a protocol essential for linear optical quantum computation, and show that it exceeds thresholds for universal cluster-state quantum computation under realistic hardware constraints. We link the device performance to the geometry of multi-photon transport, showing that localization effects from redundant, imperfect hardware may enhance robustness to coherent errors. Our design establishes a practical pathway toward near-term, scalable, and reconfigurable photonic processors in the synthetic time dimension.
\end{abstract}

\maketitle

\begin{figure*}
    \centering
    \includegraphics[width = \textwidth]{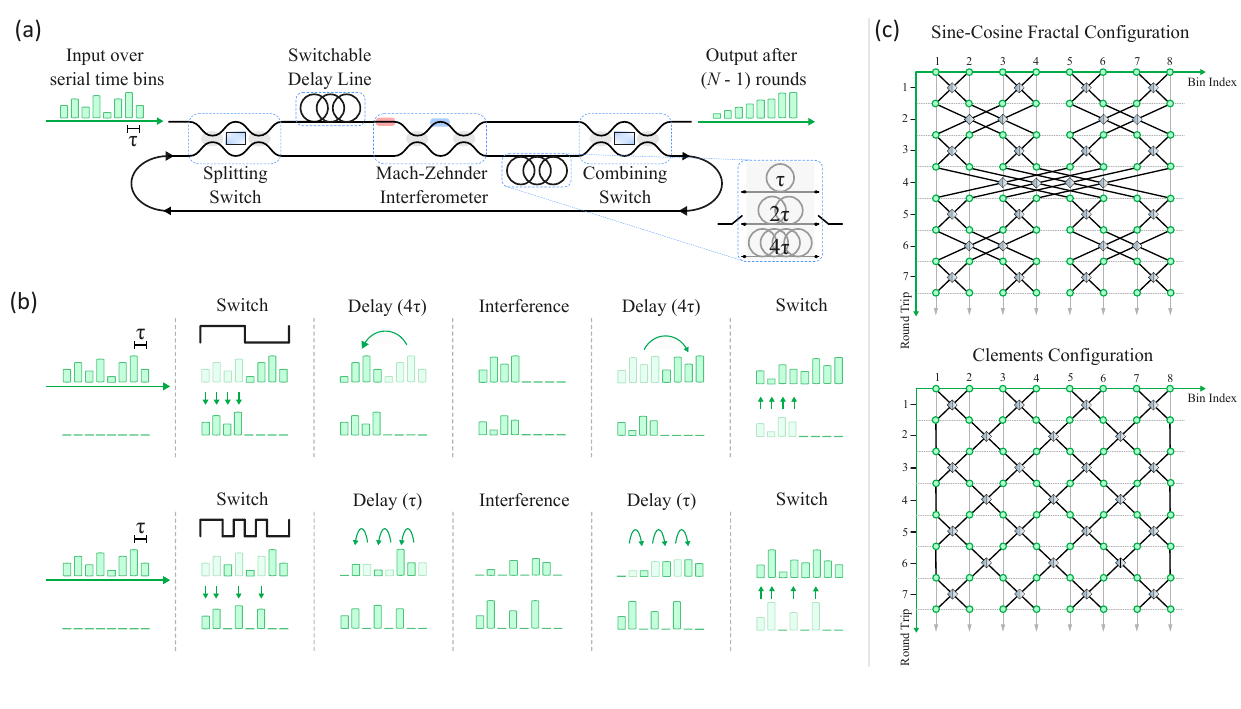}   \caption{{ Schematic and operation of the generalized Green Machine. (a) Illustration of the hardware components to construct the generalized Green Machine consisting of switches, delay lines, and programmable beamsplitters. Information is encoded serially over time bins of width $\tau$. (b) Stepwise operations indicating the processes applied to the eight time bins over two spatial modes to interfere with the fourth nearest-neighbor modes. The second delay line delays the bottom arm, equivalent to shift the time frame of the top arm to the front. (c) Fully programmable unitary transformations implemented on time-bin modes over multiple round trips in the sine-cosine fractal configuration (top) and the Clements configuration (bottom).}}
    \label{fig:schematic_main}
\end{figure*}

\section{Introduction}

Linear programmable photonic circuits with multiple inputs and outputs are fundamental building blocks throughout the development of advanced photonic processors, playing a crucial role in both classical and quantum information processing. In the classical regime, such circuits have enabled acceleration of linear algebraic computations, offering significant advantages in machine learning and AI workloads~\cite{shen2017deep, bandyopadhyay2024single, ahmed2025universal, hua2025integrated}. In the quantum domain, their capacity to implement unitary transformations among multiple modes enables high-dimensional quantum logic~\cite{luo2019quantum,zhang2019quantum,cui2020high,chi2022programmable,basani2025universal, carolan2015universal}, complex control~\cite{romero2017quantum,fischer2021autonomous,ou2025hypermultiplexed}, and precise long-range interaction of quantized photonic fields~\cite{sparrow2018simulating,qiang2018large,bao2023very}. This capability underpins quantum-advantageous protocols in photonic quantum computing, communication, and sensing, including boson sampling~\cite{madsen2022quantum, zhong2020quantum}, quantum sensing with dynamic learning~\cite{xia2021quantum,liu2025quantum}, and superadditive laser communication systems~\cite{guha2011structured,cui2025superadditive,rosati2024joint,rosati2025fourier}.

Traditionally, the most common approach to constructing programmable photonic circuits involves the use of a multiport interferometer.  This approach utilizes a mesh of beamsplitters and phase shifters to implement arbitrary unitary transformations~\cite{reck1994experimental, clements2016optimal}. Although widely demonstrated, such architectures require $\mathcal{O}(N^2)$ programmable elements to implement arbitrary unitary operations on $N$ modes, resulting in substantial hardware overhead even on integrated platforms~\cite{qiang2018large, carolan2020variational, bao2023very, sparrow2018simulating}, rendering large-scale implementations cumbersome.

In addition to scalability challenges, these individual components are highly susceptible to fabrication imperfections~\cite{bandyopadhyay2021hardware, hamerly2022accurate, hamerly2022stability, hamerly2022asymptotically}. Their individual errors accumulate as the system scales, thus introducing a challenging trade-off between scalability and achieved fidelity~\cite{burgwal2017using,kumar2021mitigating,ewaniuk2023imperfect}. The complexity is further exacerbated when extending such architectures to processors utilizing other optical degrees of freedom, such as time bins. In such cases, the implementation requires an intricate mode sorter with $\mathcal{O}(N)$ phase-stabilized optical paths, which presents formidable practical challenges in large-scale systems.

To address these challenges, alternative architectures that exploit the large-scale multiplexing capabilities of photonics have been proposed~\cite{hamerly2023multiplexing, basani2024all, ou2025hypermultiplexed, hamerly2019large}. In particular, processing information encoded within photonic time bins can significantly reduce the hardware overhead by time-multiplexing the optical components while maintaining full programmability. One popular approach~\cite{motes2014scalable,he2017time,sempere2022experimentally,monika2025quantum, yu2023universal, pegoraro2024demonstration} utilizes a pair of nested short and long optical delay lines to construct arbitrary linear transformations via the Reck-Zeilinger decomposition~\cite{reck1994experimental}. The time complexity required to compile arbitrary unitary matrices using the nested loop architecture scales as $\mathcal{O}(N^2)$. More recently, an alternative approach~\cite{bouchard2024programmable} used optically induced nonlinearities and birefringent materials to perform arbitrary unitary transformations using a synchronized pulsed pump and $\mathcal{O}(N)$ optical components in the cascaded layout.

In this manuscript, we introduce the \textit{generalized Green Machine}, a flexible time-domain photonic processor architecture for implementing programmable linear transformations on optical modes encoded in the synthetic time dimension. This recursive architecture trades complexity in the spatial domain for complexity in the temporal domain while offering enhanced flexibility. It relies on only a single Mach-Zehnder interferometer (MZI) pairing with switchable delay lines to perform interference among the time-binned modes. Unlike spatial-mode multiport interferometers, where $\mathcal{O}(N^{2})$ active MZIs create a scaling bottleneck in practice, the generalized Green Machine requires only $\mathcal{O}(\mathrm{log}_{2}N)$ to $\mathcal{O}(1)$ delay lines, depending on the adopted connectivity. Therefore, the generalized Green Machine enjoys at least an exponential reduction in the hardware component count compared to traditional spatial mode interferometer meshes.

We provide detailed prescriptions for programming the generalized Green Machine to achieve desired unitary matrices, and numerically show its robustness to coherent beamsplitter errors in the MZI with one-photon and two-photon scattering. After the end-to-end characterization, we benchmark its performance by the boosted Bell-state measurement (BSM)~\cite{ewert20143,hauser2025boosted}, a fundamental protocol in linear optical quantum computing and networking. We show that under practical hardware conditions, our proposed architecture is capable of surpassing a specific percolation threshold needed for the fusion-based generation of large-scale quantum cluster states using current photonic technology~\cite{pant2019percolation}. Owing to its flexible connectivity, the number of recursive rounds required to implement boosted BSM is significantly lower than that of the nested loop architecture restricted to the Clements decomposition~\cite{motes2014scalable}. We have further explored that these results are linked to the geometrical nature of our design in multiphoton transport and interference.

\section{Parallel interference and recursive architecture}

The layout of the generalized Green Machine is depicted in Fig.~\ref{fig:schematic_main}(a). The input photonic state is encoded over complex-valued amplitudes across serialized time bins, with a time interval of $\tau$. The MZI is parameterized by two reconfigurable phases $(\theta, \phi)$, allowing individual interference of each pair of time bins. An output port of the second switch is connected to one of the input ports, allowing photonic fields to be recursively propagated within this system or to drop out of the system. This structure enables dynamic programming of connectivity among the time-bin modes and overall circuit depth, allowing it to be configured into well-established interferometer architectures~\cite{reck1994experimental, clements2016optimal, basani2023self}.

To demonstrate the underlying working principle of the generalized Green Machine, we begin with an illustration of the interference performed in a single round trip. To achieve this function, the first $2 \times 2$ optical switch divides the input time bins into two sequences, each directed to a distinct path mode, with one sequence leading the other. The leading sequence is then delayed by an appropriate delay line to align with the lagging sequence. When the corresponding pair of time-bin modes $(i,j)$ arrives, the programmable MZI parameters, $\theta_i$ and $\phi_i$, are set to the values determined by the decomposition of the target unitary operation. After interference, two output time-bin sequences exit the MZI simultaneously. The delay line in the second arm then delays the output sequence at the bottom port, postponing it to avoid overlapping with the other sequence in time. At the end of this round, the second switch concatenates the two sequences into one. Fig.~\ref{fig:schematic_main}(b) illustrates the schematics of stepwise operations applied to 8 time-bin modes to couple fourth-nearest and nearest-neighbor modes.

Therefore, the unitary transformation for the $n^{\rm th}$ round can be represented as:
\begin{equation}
    U ^{(n)} (\vec{\theta}, \vec{\phi}) = \prod T_{i, j} (\theta_i, \phi_i),
\end{equation}
where $T_{i, j}$ is the $2\times 2$ unitary matrix among selected pairs of time bins $i$ and $j$, parametrized as:
\begin{equation}
    T_{i,j}(\theta,\phi) = i e^{i\theta/2} \begin{bmatrix} e^{i\phi} \sin(\theta/2) & \cos(\theta/2) \\ e^{i\phi} \cos(\theta/2) & -\sin(\theta/2) \end{bmatrix}. \label{eq:mzi_t}
\end{equation}

The output state, after undergoing the unitary transformation $U^{(n)}$ in the $n^{\rm th}$ round, is fed back into the apparatus for the next round. This process is repeated until the desired global unitary transformation is performed, and the second switch can be programmed to send the combined output sequence to the drop-off port.

By compiling interference patterns in time (via multiple programmed recursive rounds), the generalized Green Machine can be configured to perform arbitrary unitary transformations. The sine-cosine fractal (SCF) architecture~\cite{basani2023self}, shown in Fig.~\ref{fig:schematic_main}(c), is composed of stages with $2^{k}$ nearest-neighbor connectivity, where $k \in [0, 1, 2, \hdots, \mathrm{log}_{2}N-1]$. To realize the SCF mesh architecture, $\mathrm{log}_{2}N$ delay lines of length $2^{k}\tau$ are sufficient. Alternatively, the device can also be programmed into the Clements architecture~\cite{clements2016optimal}
by programming alternate stages to couple the nearest-neighbor modes. In this case, a \textit{single} delay line of length $\tau$ is sufficient. The stepwise operations, indicating the processes applied to the time-bin modes to configure them into Clements and SCF configurations, are detailed in Appendix A. 

\section{Performance under Imperfections}

\begin{figure}
    \centering
    \includegraphics[width = \columnwidth]{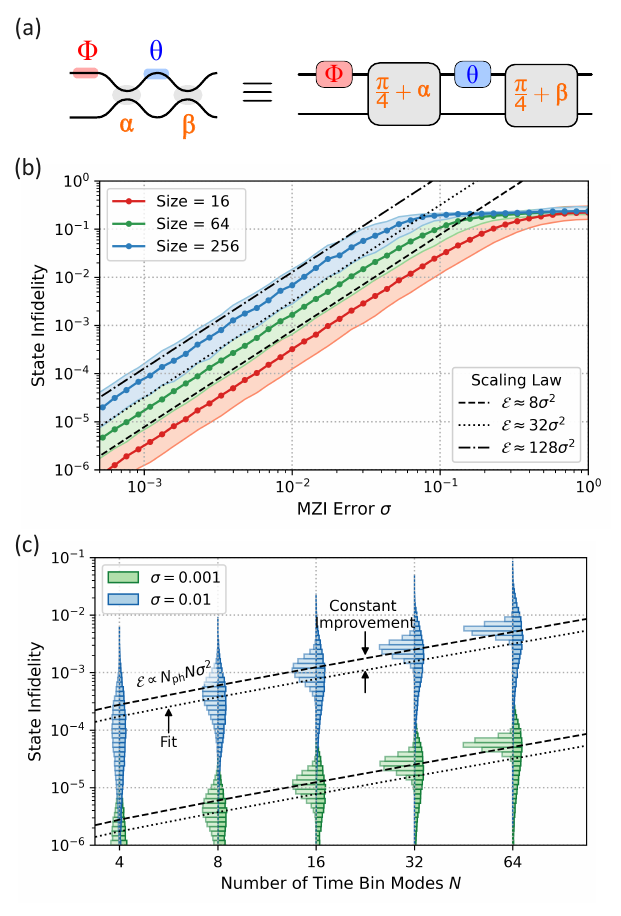}
    \caption{{Performances of the generalized Green Machine under beamsplitter errors. (a) Model of MZI with coherent errors $(\alpha, \beta)$ causing deviations from the ideal 50:50 splitting. (b) State infidelity as a function of beamsplitter errors $\sigma$ for a single-photon state scattered among $N = 256$ time bins. The simulated performance outperforms the dashed lines that mark the scaling law of traditional spatial-mode interferometers. (c) Histograms of the infidelities of the two-photon boson sampling among up to  $N = 32$ time bins. Infidelities due to the correlated model of errors from the generalized Green Machine are shown in the darker color (right half), fitted with the dotted line. The brighter (left half) histograms illustrate the distribution of infidelities under uncorrelated errors (as seen in traditional multiport interferometer architectures) in since-cosine fractal mesh, and are fitted with the dashed line. }}
    \label{fig:GM_performance}
\end{figure}

Errors in the generalized Green Machine stem from two main sources. The first arises from component imperfections, which cause the beamsplitters in the MZI to deviate from their ideal 50:50 splitting ratio~\cite{bandyopadhyay2021hardware, hamerly2022accurate, hamerly2022stability, vadlamani2023transferable, hamerly2022asymptotically}. The second is the imbalanced insertion loss introduced by the switchable delay modules and MZI. We discuss the impact of coherent errors in this section, and the impact of loss in Appendix C. 

Imperfect splitting in the beamsplitter ratios, caused by variations in the fabrication process, introduces errors in the programmed unitary matrix. Unlike spatial-mode interferometers, the generalized Green Machine utilizes only a single MZI, resulting in correlated errors among all the time-bin modes. Under the influence of these correlated errors, the transfer matrix implemented by the generalized Green Machine $U_{\mathrm{GM}}$ deviates from the ideal targeted unitary matrix $U_{\mathrm{ideal}}$. To quantify the impact of these coherent errors, we evaluate the average state infidelity of a group of few-photon states $\ket{\psi}$ that evolve through $U_{\mathrm{GM}}$ as: 
\begin{equation}
\bar{\mathcal{E}} = 1 - \bar{\mathcal{F}} = 1 - \langle~ \, |\langle \psi | U_{\mathrm{ideal}}^\dagger U_{\mathrm{GM}} | \psi \rangle|^2~\rangle
\end{equation}

The impact of component imperfections on the Green Machine’s ability to implement large-scale unitaries is benchmarked by simulating the scattering of a single photon across $N \in  [16, 64, 256]$ time bins. For a given $N$, we sample $10^{4}$ unitaries from the Haar measure, each with a random input state. The beamsplitter errors for each circuit, $(\alpha, \beta)$, are drawn independently from the Gaussian distribution $\mathcal{N}(0, \sigma)$. Figure~\ref{fig:GM_performance}(a) illustrates a schematic of an MZI with the two beamsplitters perturbed by component errors $(\alpha, \beta)$. In Fig.~\ref{fig:GM_performance}(b), we plot the state infidelity $\mathcal{E}$ as a function of beamsplitter error $\sigma$. The median and interquartile ranges (colored bands) are plotted alongside analytical predictions for spatial-mode interferometers with uncorrelated errors, which scale as $\mathcal{E} \approx \frac{1}{2}N\sigma^{2}$  (black lines. See Appendix B for the analytical derivation). 

Then, to explore its potential in multiphoton transport and boson sampling, we benchmark the generalized Green Machine on sampling an $N_{\mathrm{ph}} = 2$ photon state from random unitaries of up to size $N = 64$. Fig.~\ref{fig:GM_performance}(c) compares distributions of sampling infidelities for the Green Machine (right half, darker) and a conventional spatially meshed multiport interferometer (left half, brighter) with uncorrelated errors under two noise levels: $\sigma = 0.001$ plotted in green and $\sigma = 0.01$ plotted in blue. The dotted and dashed lines are fitted to the median values of these distributions, and follow the scaling law:
\begin{equation}
    \mathcal{E} \propto N_{\mathrm{ph}}N\sigma^{2}.
\end{equation}
As shown in Fig.~\ref{fig:GM_performance}(b) and \ref{fig:GM_performance}(c), for both cases, the generalized Green Machine reduces the mean infidelity by a factor of $\sim\sqrt{2}$ across all system sizes by virtue of its correlated error model (discussed in Appendix B). Despite a marginally larger variance, the resulting broader distribution increases the likelihood of postselecting high-performance hardware. Notably, the analytical scaling of spatial-mode interferometer meshes under uncorrelated coherent errors remains identical, regardless of whether the decomposition follows the Clements, Reck-Zeilinger, or SCF architecture.~\cite{hamerly2022accurate, hamerly2022stability}. In practice, physical implementations of the generalized Green Machine may experience a combination of both correlated and uncorrelated errors. Uncorrelated noise may arise from fluctuations in control signals or the thermal drift of the MZIs. If this noise is uncorrectable, the fidelity of the generalized Green Machine will typically fall between the optimal correlated-noise limit (with a $\sqrt{2}$ advantage) and the standard baseline performance of a spatial mesh with completely uncorrelated noise, depending on the relative dominance of these error sources.

While Fig.~\ref{fig:GM_performance}(c) compares the infidelity of spatial-mode interferometer meshes (uncorrelated errors) with that of the generalized Green Machine (correlated errors), an additional advantage of the Green Machine emerges when hardware error-correction techniques are applied~\cite{bandyopadhyay2021hardware, basani2023self}. These techniques further improve the scaling of corrected matrix error from $\mathcal{O}(N\sigma^2)$ to $\mathcal{O}(\sqrt{N\mathrm{log}_{2}N}\sigma^2)$
providing a substantial reduction in coherent-error accumulation. Fig.~\ref{fig:app_corrected_scaling} provides a direct comparison of matrix errors across all four regimes: spatial-mode interferometers with uncorrelated errors, the generalized Green Machine with correlated errors, and both schematics after hardware error correction.

\section{Toward Robust Boosted Bell-State Measurement}

\begin{figure}
    \centering
    \includegraphics[width = \columnwidth]{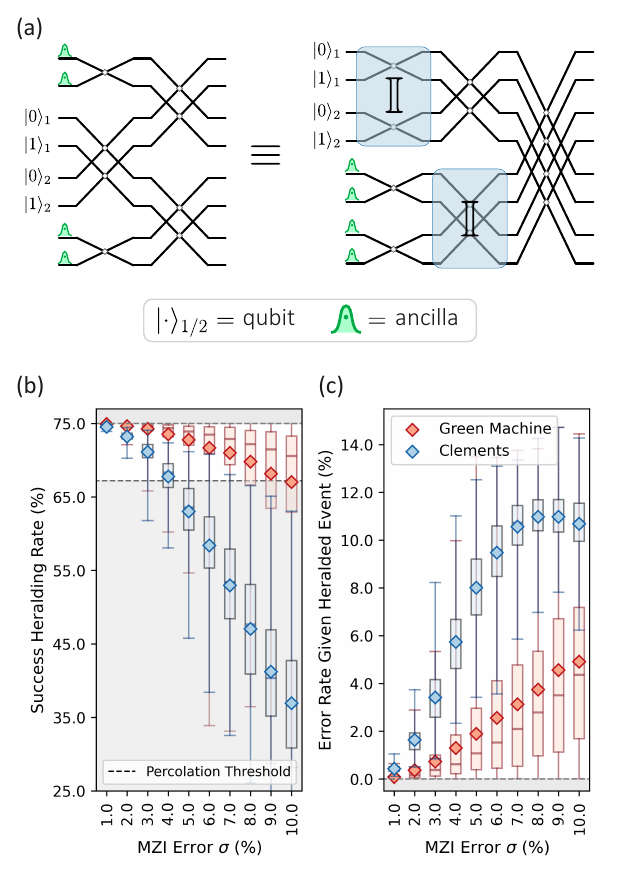}    \caption{ Green Machine implementation of boosted Bell-state measurement. (a) Circuit of the boosted Bell-state measurement implemented by a minimum of 3 stages of the Green Machine on two dual rail qubits and four ancillary modes to achieve a successful prediction probability of 75\%. The blue shaded regions indicate sections of the Green Machine circuit set to the identity operation. (b) Probability of successful prediction as a function of MZI coherent-error characteristics $\sigma$ comparing the performance of the 3-stage Green Machine with the 7-stage Clements mesh. (c) Error given successful prediction as a function of MZI coherent-error characteristics $\sigma$.}
    \label{fig:GM_BBM}
\end{figure}

A near-term application that immediately benefits from our architecture is the boosted BSM~\cite{ewert20143}. The BSM aims to project a dual-rail-encoded photon pair onto four Bell states: $|\Psi_{\pm}\rangle = \frac{|01\rangle \pm |10\rangle}{\sqrt{2}}$ and $|\Phi_{\pm}\rangle = \frac{|00\rangle \pm |11\rangle}{\sqrt{2}}$, which is a fundamental operation for the fusion-based generation of a large-scale cluster state~\cite{pant2019percolation}, quantum teleportation, and entanglement swapping~\cite{dhara2023entangling}. The standard BSM circuit with a simple beamsplitter operation can only deterministically distinguish between $|\Psi_{\pm}\rangle$ from measurement results, which places an upper bound on the success rate at 50\%~\cite{calsamiglia2001maximum}. This success rate can be boosted by incorporating ancillary single-photon modes~\cite{ewert20143} and a larger multiport interferometer. An example of such a circuit and its equivalent implementation using three stages of the generalized Green Machine is shown in Fig.~\ref{fig:GM_BBM}(a). The positions of the dual rail qubits are denoted by $\ket{\cdot}_{1/2}$ while the ancillary modes are denoted by rails populated with a single photon, indicated by the green pulse. This circuit increases the BSM success rate to 75\% in ideal cases--surpassing the percolation threshold of 67.2\% necessary for universal photonic quantum computing with cluster states~\cite{pant2019percolation}, and increases the efficiency of entanglement generation for quantum networks. An added advantage of using the time-bin architecture is that the ancillary single photons can be sequentially generated from a selected single-photon emitter, ensuring consistent high indistinguishability necessary for high-fidelity quantum interference.

The performance of boosted BSM is characterized by two figures of merit: (1) the \textit{success heralding rate}, defined as the probability of heralding the projection into one Bell state according to its detection signature, and (2) the \textit{error rate given heralded event}, which quantifies the misidentification rate due to measurement crosstalk after heralding. We employ a Bayesian inference model that assigns posterior probabilities to each detection signature, distinguishing between two outcomes: 
\begin{itemize}
    \item \textit{Decode}, where the event projects the received pattern onto one of the four Bell states if the posterior probability is greater than the decision threshold. This contributes to the success heralding rate. A nonzero probability of detecting the same click patterns by the other Bell states contributes to the increased error rate given this successful prediction.
    \item \textit{Discard}, where no assignment is made if the posteriors are inconclusive. This results from an inability to decide among all four Bell states, and reduces the heralding success rate.
\end{itemize}

The success rate predicted by the Bayesian model is subject to variation due to several factors. These include sources of crosstalk in the time-domain boosted BSM circuit, such as coherent beamsplitter errors or tunable parameters like circuit depth and decision threshold. Our results are plotted over 1000 samples taken from circuits with emulated noises, as quantile box plots with the mean value indicated by the red diamond.

\begin{figure}
    \centering
    \includegraphics[width = \columnwidth]{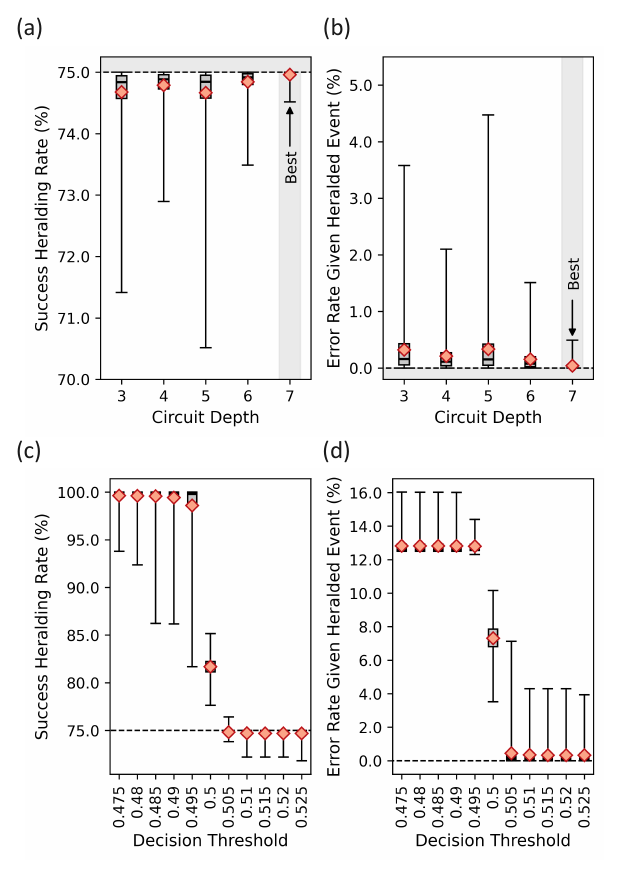} \caption{{ Successful heralding rate and error rate given heralded event for boosted Bell-state measurement. Probability of successful heralding and error given heralded event as a function of: (a, b) circuit depth, varying from maximally pruned to fully expressive (c, d) decision threshold with 2\% MZI error.}}
    \label{fig:GM_BBM_sweeps}
\end{figure}

First, we evaluate the impact of coherent errors by sampling the errors $(\alpha, \beta)$ from a normal distribution $\mathcal{N}(0, \sigma)$. Results plotted in Fig.~\ref{fig:GM_BBM}(b) and~\ref{fig:GM_BBM}(c) compare the successful prediction rate (and a corresponding increase in the error rate) for the boosted BSM circuit implemented using the Green Machine and the Clements mesh. For the Green Machine (indicated in red), at a splitter error of $\sigma = 3\%$, the worst-case success rate drops below the percolation threshold. The average success rate drops to this threshold at $\sigma \sim 10\%$, where more than a quarter of the sampled circuits have their heralding success rates below the threshold. In contrast, an eight-mode spatial Clements interferometer mesh requires the full circuit depth of all seven stages to implement the equivalent matrix transformation. More stages accumulate more uncorrelated MZI error, further degrading the fidelity of the boosted BSM circuit. As shown in Fig.~\ref{fig:GM_BBM}(b) and \ref{fig:GM_BBM}(c), the average success rate of the spatial-domain Clements implementation falls to the percolation threshold at an error rate of $\sigma\sim 4\%$, whereas the average success rate of the Green Machine hits the threshold at an error rate of $\sigma\sim 10\%$. For the boosted BSM error rates given heralded events, the Clements implementation is also more sensitive to the MZI error than the Green Machine. 

Subsequently, we vary the circuit depth by adding redundant stages in the configuration of the SCF mesh. Contrary to expectations that deeper, noisier circuits degrade performance, we observe in Fig.~\ref{fig:GM_BBM_sweeps}(a) and (b) a nonmonotonic improvement in both success and error rates. This counterintuitive behavior is linked to multiphoton transport phenomena, which we analyze in the following section. In practice, the three-stage Green Machine may still be the best option, considering the additional loss brought by redundant stages.

Finally, we sweep the decision threshold, which is used to determine the decoding strategy employed in our inference model. By reducing this threshold, the successful prediction rate can increase at the expense of a higher error rate, as shown in Fig.~\ref{fig:GM_BBM_sweeps}(c) and~\ref{fig:GM_BBM_sweeps}(d). With $\sigma=2\%$ MZI error, the transition edge between inference strategies becomes less sharp, which can serve as a tuning knob for globally optimizing the overall performance of fusion-based photonic quantum computing in practice. 

\begin{figure*}
    \centering
    \includegraphics[width = \textwidth]{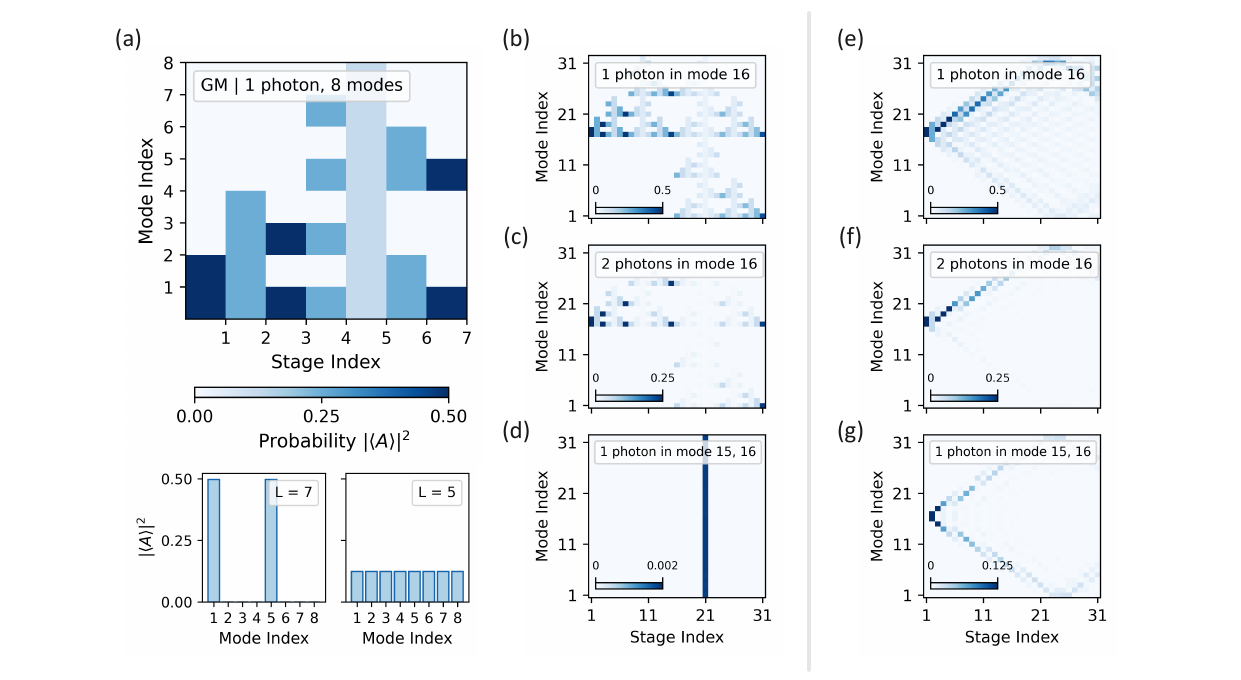}  \caption{{ Photon transport through the generalized Green Machine. (a) Single photon transport through the 8-mode circuit with all MZIs setting to 50:50, showing a localized state at $L = 7$, and a diffused state at $L = 5$. (b) Single photon transport with varying degrees of localization in a 32-mode circuit. The heatmap indicates the probability of measuring a photon at each mode after each stage. (c) and (d) Two-photon transport depicting varying degrees of localization in the presence of Hong-Ou-Mandel interference, indicating that the localization and diffusion still happen with quantum interferences. The heatmap indicates the probability of two-photon meeting at the same mode after each stage. (e) Single photon transport in a 32-mode Clements configured circuit, showing no distinct regimes of transport. (f, g) Two-photon transport statistics (probability of two-photon meeting at the same mode) through the Clements configuration.}}
    \label{fig:GM_transport}
\end{figure*}

\section{Self-Similar Quantum Transport}

The performance of the generalized Green Machine is strongly linked to the geometrical nature of quantum transport through modes~\cite{harris2017quantum}. An example of this behavior is distinctly visible in our simulations of the boosted BSM in Fig.~\ref{fig:GM_BBM_sweeps}(a), where the success rate drops at a circuit depth of 5, and increases significantly at a depth of 7. To explain this phenomenon, we study photon transport through the generalized Green Machine with it configured to the SCF mesh.

We model this transport by tracking the photon statistics of single and two-photon states hopping from one input time-bin mode to other modes after each round averaged over 100 noisy circuits sampled with $\sigma=2\%$. An initial state of one or two photons (either a two-photon Fock state or two unentangled indistinguishable photons) is used as the input in our simulations. This optical state interferes across the time-binned modes as it evolves through the stages, with an MZI that is always set to 50:50 splitting ratio. These amplitudes are plotted in Fig.~\ref{fig:GM_transport} for 8 (similar to the boosted BSM circuit) and 32 time-bin modes. The versatility of the generalized Green Machine enables us to contrast the transport dynamics under different connectivity configurations. This fundamental difference in transport dynamics observed between the SCF and Clements configurations is an expected consequence of their internal connectivity and unit-cell structures. The SCF configuration possesses a fractal, self-similar structure that leads to complex and nonmonotonic transport patterns, arising from re-interference. On the other hand, the Clements architecture interferes only nearest-neighbor modes which results in systematic diffused mode-mixing.

These simulation results reveal distinct, visually structured regimes of transport, ranging from fully diffused to strongly localized in the SCF configuration. For the case of single-photon transport in an 8-mode system, illustrated in Fig.~\ref{fig:GM_transport}(a), we observe a fully diffused regime at stage 5, where the probability of detecting the photon is uniformly distributed across all modes. In contrast, at stage 7, the optical state becomes highly localized, with significant amplitude confined to just two modes. This alternating behavior between delocalized and localized regimes persists in larger systems, such as the 32-mode configuration shown in Fig.~\ref{fig:GM_transport}(b), where the transport patterns reflect the self-similar structure of the SCF mesh.

The introduction of disorder--particularly static perturbations arising from beamsplitter imperfections--modifies these transport characteristics. This disorder has been shown to cause coherent particles to become exponentially localized--a phenomenon commonly referred to Anderson localization~\cite{anderson1958absence, harris2017quantum}. In our system, this would correspond to the exponential localization of the optical mode to its respective time bin. We attribute the nonmonotonic behavior of the boosted BSM success rate in Fig.~\ref{fig:GM_BBM_sweeps}(a) and (b) to such localization effects. Specifically, the robustness observed at stage 7 may stem from localization-induced protection, in contrast to the fragility of the diffused state at stage 5, as shown in Fig.~\ref{fig:GM_transport}(a). The impact of these localization effects and their ability to protect quantum information from coherent errors remains the subject of our future work.

The effects resulting from this self-similarity also produce distinct regimes in the case of multiphoton transport. As an example, we simulate two-photon transport in a 32-mode system, shown in Figs.~\ref{fig:GM_transport}(c) and (d). The resulting interference pattern exhibits a similar fractal pattern, with populations alternating between the single and two-photon excitation manifolds, when the input state is a two-photon Fock state (see Fig.~\ref{fig:GM_transport}(c)). In another example, when modes 15 and 16 are excited using identical single photons, only at $21^{\mathrm{st}}$ stage do we see two photons meeting at the same mode, where the probability is distributed uniformly over all modes, as shown in Fig.~\ref{fig:GM_transport}(d). These are in sharp contrast to both single- and two-photon transport across the Clements configuration with identical input states, as shown in Fig.~\ref{fig:GM_transport}(e), (f), and (g), where the photon evolution trajectories diffuses from the input modes.

\begin{table*}
    \centering
    {\begin{tabular}{c|c|c|c|c}
    \hline \hline 
     Architecture & ~\makecell{Hardware \\ Complexity}~ & ~\makecell{Throughput \\ density}~ & ~\makecell{Compilation \\ time}~ & ~Loss scaling \\ \hline
      
      This work (Clements)  & $\mathcal{O}(1)$ & $\mathcal{O}(N)$ & $\mathcal{O}(N^{2}\tau)$ & $\mathcal{O}(N(\eta_{\mathrm{bs}} + \eta_{\mathrm{i}} + c\tau\eta_{\mathrm{o}}N))$\\
     
     This work (SCF)  & $\mathcal{O}(\mathrm{log}_{2}N)$ & $\mathcal{O}(N/\mathrm{log}_{2}N)$ & $\mathcal{O}(N^{2}\tau)$ & $\mathcal{O}(N(\eta_{\mathrm{bs}} + c\tau\eta_{\mathrm{i}}\mathrm{log}_{2}N + c\tau\eta_{\mathrm{o}}N))$\\
     
     Clements (spatial)~\cite{clements2016optimal} & $\mathcal{O}(N^{2})$ & $\mathcal{O}(1)$ & $\mathcal{O}(N)$ & $\mathcal{O}(N\eta_{\mathrm{bs}})$\\

     Motes \textit{et al.}~\cite{motes2014scalable} & $\mathcal{O}(1)$ & $\mathcal{O}(N/2)^{\ddagger}$ & $\mathcal{O}(N^{2}\tau)$ & $\mathcal{O}(N(\eta_{\mathrm{bs}} + \eta_{\mathrm{i}} + c\tau\eta_{\mathrm{o}}N))$ \\

     Bouchard \textit{et al.}~\cite{bouchard2024programmable} & $\mathcal{O}(N)$ & $\mathcal{O}(N)$ & $\mathcal{O}(N\tau)$ & $\mathcal{O}(N(\eta_{\mathrm{bs}}+\eta_{\mathrm{i}}))$\\
     
    \hline \hline
    \end{tabular}}
    \caption{ Comparison of performance metrics for approaches to unitary transformations on optical modes. Throughput density is defined in units of mulitply accumulate operations per second per pass per hardware. $\ddagger$ We emphasize that the throughput of Motes' architecture is half of this work. Here $\tau$ denotes the temporal spacing of neighbor time bins (including bin size and guard band if exists), $\eta_{\mathrm{bs}}$ denotes the loss in dB per Mach-Zehnder interferometer, while $\eta_{\mathrm{i}}$ and $\eta_{\mathrm{o}}$ denote the loss rate in dB/m for the inner and outer delay lines respectively, and $c$ stands for the speed of light in the delay lines. We note that the architecture of Bouchard et al. \cite{bouchard2024programmable} can be naturally generalized into the broader operational framework of the Green Machine architecture.}
    \label{tab:comparision}
\end{table*}

\section{Discussion and Outlook}

We have presented an architecture for the \textit{generalized Green Machine}, a hardware-efficient, universal time-domain linear optical processor that utilizes dual switches. This architecture is naturally compatible with both Clements~\cite{clements2016optimal}, SCF mesh~\cite{basani2023self}, or a hybrid of both, to express arbitrary linear unitary matrices in distinct symmetries with a minimum number of round trips. We have numerically simulated its performance under practical imperfections in the task of boosted BSM, single-photon, and two-photon transport, in which our new architecture gains a constant scaling advantage in average fidelity compared to spatial-mode multiport interferometers. Other than discrete-variable quantum photonics, the proposed architecture also works for continuous-variable quantum photonics as it operates without unintended vacuum input modes, which leads to new opportunities for all-photonic generation of cat and GKP states~\cite{su2019conversion}, quantum computing~\cite{konno2024logical,aghaee2025scaling}, and new practical quantum photonic applications~\cite{guha2025quantum} focusing on time-domain information processing.

The two leading figures of merit that can be used to benchmark across linear optical processors are (1) the hardware complexity, defined as the number of hardware components required to implement an arbitrary $N \times N$ unitary matrix, and (2) the throughput density, defined as the number of multiply accumulate (MAC) operations per round-trip per amount of hardware required. In terms of hardware complexity, we require at best $\mathcal{O}(1)$ components in the Clements configuration or $\mathcal{O}(\mathrm{log}_{2}N)$ components in the SCF configuration, providing at least an exponential reduction in the amount of hardware. Since our device operates on all $N$ modes in a single round trip, we achieve a linear scaling in throughput density, which is typically realized only with hyper-multiplexing. The comparison with other competing architectures is exhibited in Table~\ref{tab:comparision}. An additional advantage of the generalized Green Machine is that, since it can be programmed into well-understood interferometric configurations, it is amenable to self-configuration techniques~\cite{hamerly2022accurate, hamerly2022stability, basani2023self}, making it more robust to hardware imperfections.

The generalized Green Machine also features flexibility as it allows for reducing the depth of the linear optical circuit by pruning~\cite{basani2023self, yu2023heavy}--simply by reducing the number of recursive rounds. On the contrary, spatial-mode linear processors cannot be dynamically pruned--this would involve either bypassing the redundant hardware or physically detaching it. By programming the pruning scheme, the all-to-all connectivity can be maintained while minimizing loss and latency. This is particularly beneficial for implementing classes of transformations with specific symmetries. For instance, achieving an eight-mode boosted BSM circuit requires only three recursive rounds, while achieving an $N\times N$ Hadamard transform requires only $\mathrm{log}_{2}N$ recursive rounds~\cite{guha2011structured}. The proof-of-concept demonstration of a 16-mode Hadamard transform in a fiber-optical setup has been shown in Ref.~\cite{cui2025superadditive}. 

This capability to dynamically program the connectivity, depth, and drop-out ports of the circuit enables the generalized Green Machine to be programmed into more general (i.e., nonunitary) beamsplitter meshes~\cite{hamerly2025toward} such as the diamond~\cite{mosca2002novel, taballione20198} or path-independent loss (PILOSS) architectures~\cite{suzuki2014ultra, suzuki2019low}. Large-scale and high-dimensional transformations can be performed by spatially multiplexing our device. We discuss this hypermultiplexed architecture in Appendix D and evaluate its computational speed for accelerating matrix products in classical machine learning tasks.

The loss of the programmable linear optical circuit constitutes another critical figure of merit. Quantum applications relying on multiphoton transport usually experience a severe decline in performance due to such loss. Here, we analyze the loss tolerance for boosted BSM against a baseline success heralding rate exceeding the percolation threshold of 67.2\%. In the worst-case scenario--postselecting for the detection of all six photons, the success heralding rate scales as $P = 0.75 \eta^6$ with $\eta$ the total transmission of the circuit. This imposes a minimum transmission requirement of $\eta \geq 0.98$ (loss $\leq$ 0.08 dB), which is achievable with near-term integrated photonic technologies~\cite {he2019low,chang2017heterogeneous, psiquantum2025manufacturable}, nonlinear optics~\cite{bouchard2024programmable}, or using free-space optics~\cite{madsen2022quantum,arnold2023free}. Notably, estimates from Refs.~\cite{melkozerov2024analysis, maring2024versatile} suggest that this bound may be further relaxed where losing photons does not destroy all the information. In contrast, a spatial-mode Clements interferometer requires a depth of seven layers to realize the same transformation, which exacerbates the drop in success rate and the rise in error rates given heralded event, rendering practical implementation significantly more challenging.

Similarly, optical loss compromises the quantum advantage of boson sampling by rendering the sampling problem classically tractable~\cite{oh2024classical}. Our proposal’s dynamic connectivity, however, allows us to realize Haar-random matrices with the minimum necessary optical path length through well-established optimization routines. This ability to dynamically prune circuits directly reduces total optical depth and accumulated loss while perfectly preserving the all-to-all connectivity required for universality--a capability typically unavailable to spatial-mode interferometer meshes.

In practice, the optical loss of the generalized Green Machine stems primarily from fast optical switches and optical delay lines. As detailed in Table~\ref{tab:comparision}, the recursive architecture introduces additional loss through $\mathcal{O}(N)$ passes of the outer-loop delay line, scaling as $c\tau\eta_{\mathrm{o}}\mathcal{O}(N^2)$. However, for a system with $N=100$ optical modes and a time-bin width of $\tau=100$ ps, this accumulated loss is approximately 0.04 dB using single-mode fiber at 1550 nm ($\approx 200$ m with a loss rate of 0.2 dB/km). This is negligible compared to the measured 0.1 dB per MZI facet made with thin-film Barium Titanate BTO~\cite{psiquantum2025manufacturable}. Therefore, using present-day hardware, our architecture offers loss figures comparable to--or, with layer pruning, even lower than those of state-of-the-art fast-programmable spatial interferometer meshes.

The generalized Green Machine achieves a significantly reduced hardware footprint at the cost of temporal overhead, with its compilation time scaling as $O(N^2 \tau)$. This contrasts with spatial-mode interferometer meshes, where compilation is time-of-flight limited and scales as $O(N)$. However, the $O(N^2 \tau)$ scaling for the generalized Green Machine represents a worst-case upper bound. While in practice, circuits can be dynamically pruned to minimize the round trips required for a specific target unitary.

To put this overhead into perspective, we consider a system of size $N=100$ utilizing state-of-the-art optical switches ($\tau \approx 4.3$ ps)~\cite{bouchard2024programmable}. The latency to compile an arbitrary unitary is approximately 43 ns, which is comparable to the detection and data acquisition times of a spatial-domain processor. Nevertheless, for extremely large-scale systems where quadratic scaling overtakes the switching speed (e.g., $N^{2} > 1~\text{sec}/\tau$ at $N\sim 10^{5}$), this compilation time could become the dominant constraint on throughput.

\section*{Acknowledgments}
All authors acknowledge the DARPA PhENOM program for the support. C.C. and S.G. also acknowledge the DARPA QuANET program for partial support, and thank the support from the Engineering Research Center for Quantum Networks (CQN) under NSF Grant No. EEC-1941583 for synergistic research support. 

\section*{Data Availability}
The data that support the findings of this article are openly available~\cite{Basani2024_CasOptAx}.

\section*{Appendix A: Compiling Temporal Mode Transformations \label{sec:appendix_1}}

\begin{figure}[h!]
    \centering
    \includegraphics[width =\columnwidth]{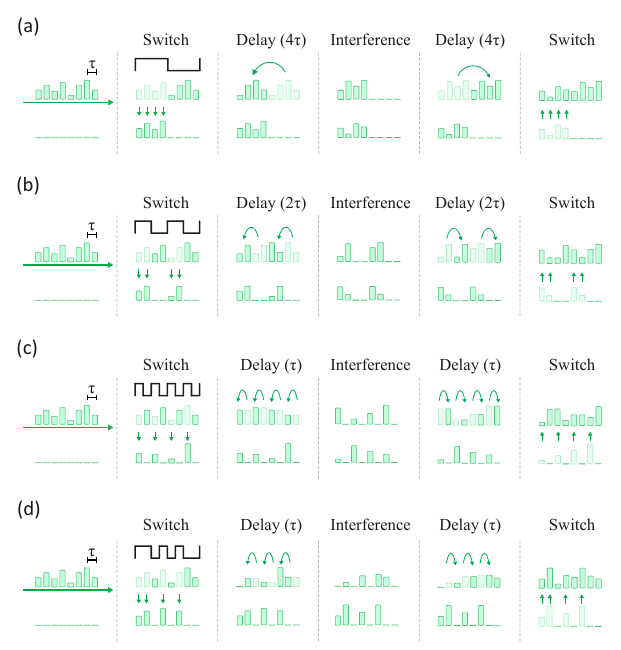}
    \caption{{ Stepwise operations to couple $n^{\mathrm{th}}$ nearest time-bin modes. Illustration of the switching scheme and the impact of interference on eight time-bin modes. These steps can be used to compile global unitary operations via the SCF configuration using (a), (b), and (c) or the Clements configuration via (c) and (d).}}
    \label{fig:app_switching}
\end{figure}

The gGM architecture we introduce in the main text compiles transformations stage-wise on time-bin modes to perform a desired unitary operation. Depending on the preprogrammed configuration being used (Clements, SCF, or otherwise), each stage couples the $n^{\mathrm{th}}$ nearest-neighbor modes. In Fig.~\ref{fig:app_switching} , we illustrate the stepwise operations, indicating the processes being applied to eight time-bin modes. Operations demonstrated in  Fig.~\ref{fig:app_switching}(a), (b), and (c) are sufficient to realize arbitrary 8-mode unitaries in the SCF configuration by coupling fourth-nearest, second-nearest, and nearest-neighbor modes. Similarly, operations in Fig.~\ref{fig:app_switching}(c) and (d) are sufficient to realize the Clements configuration by coupling odd and even nearest neighbor modes.

\begin{figure*}
    \centering
    \includegraphics[width = \textwidth]{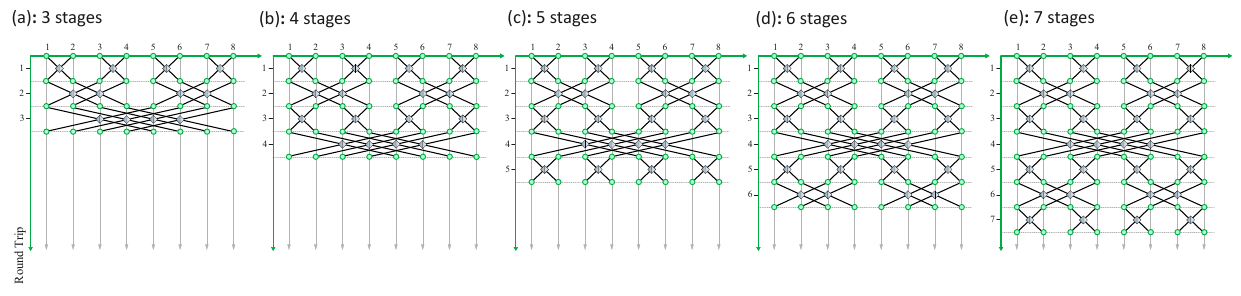}
    \caption{{ Connectivity as a function of number of stages. (a)-(e) Schematics of the connectivity among eight time-binned modes implementing different circuit depths from three to seven stages as performed in our analysis of the boosted Bell-state measurement.}}
    \label{fig:app_pruned_order}
\end{figure*}

These stage-wise interference patterns can be compiled to realize a fully expressive unitary, or pruned meshes that cannot express the entire $\texttt{SU(N)}$ unitary group. In our analysis of the boosted Bell-state measurement, we interpolate between the circuit shown in Fig.~3(a) of the main text and the fully expressive SCF mesh following the scheme proposed in ref.~\cite{basani2023self}. Figure~\ref{fig:app_pruned_order} illustrates the order in which interference patterns are compiled.

\section*{Appendix B: Derivation of Error Scaling \label{sec:appendix_2}}

We are interested in the infidelity of the output state introduced by coherent errors arising from beamsplitter imperfections in unitaries implemented by the generalized Green Machine. Consider a single-photon state $\ket{\psi} = \sum c_{i}\hat{a}^{\dagger}_{i} \ket{0}$ that evolves under a unitary implemented by the Green Machine. The notation $\hat{a}_{i} \left( \hat{a}_{i}^{\dagger} \right)$ is the annihilation (creation) operator for a bosonic excitation in the $i^{\mathrm{th}}$ time bin, and $c_{i}$ is its corresponding probability amplitude such that $\sum |c_{i}|^{2} = 1$.

We quantify the infidelity of the output state introduced by small coherent errors arising from beamsplitter imperfections. We consider a quantum state $\ket{\psi}$ evolving under a $N\times N$ unitary transformation $U$. Due to fabrication imperfections, the implemented unitary deviates from the target ideal unitary $U_{\mathrm{ideal}}$ as $U = U_{\mathrm{ideal}} + \Delta U$. By averaging the state infidelity over a group of random input states, the infidelity $\bar{\mathcal{E}} = 1 - \bar{\mathcal{F}}$ is dominated to the second order by the variance of the error operator:
\begin{equation}
    \bar{\mathcal{E}} \approx  \frac{1}{4N}||\Delta U||^2 = \frac{1}{4N}\mathrm{Tr}(\Delta U^\dagger \Delta U)
\end{equation}
where $||\cdot||$ denotes the Frobenius norm. The coefficient $1/4N$ normalizes the infidelity to $[0,1]$, matching the definition of the average state infidelity. This total error $\Delta U$ is accumulated by all the imperfections of every MZI. To analyze the scaling of $\Delta U$, we first determine the form of the perturbation $\Delta U_\ell$ at a single MZI.

A general physical MZI is parametrized by two angles/phases $\phi$ and $\theta$. $\theta$ determines the splitting ratio which is susceptible to beamsplitter coherent errors $\alpha$ and $\beta$ (coherent errors that affect $\phi$ can be compensated by calibration). We model this noisy $l^{\rm th}$ MZI (2-by-2) unitary $U_l(\theta_l, \alpha_l, \beta_l) = e^{-i (\beta_l+\pi/4) \sigma_x} e^{-i \frac{\theta_l}{2} \sigma_z} e^{-i (\alpha_l+\pi/4) \sigma_x}$ with the target splitting angle $\theta_l$, errors $\alpha_l, \beta_l$ sampled from $\mathcal{N}(0, \sigma)$, and $\sigma_{x,y,z}$ representing 2-by-2 Pauli matrices. Expanding around the target ideal unitary $U_0$, the first-order error perturbation (for small $\alpha_l$ and $\beta_l$) is:

\begin{equation}
\begin{split}
\Delta U_\ell \approx & - \left[ (\alpha_\ell+\beta_\ell)\cos\frac{\theta_\ell}{2} + i(\alpha_\ell-\beta_\ell) \sin\frac{\theta_\ell}{2}\sigma_y \right] \\
& + \mathcal{O}(\alpha_\ell^2, \beta_\ell^2, \alpha_\ell\beta_\ell).
\end{split}
\end{equation}
Now, combining the independent errors into symmetric ($S_\ell = \alpha_\ell + \beta_\ell$) and antisymmetric ($A_\ell = \alpha_\ell - \beta_\ell$) error components, it yields:
\begin{equation}
    \Delta U_\ell \approx - \left[ S_\ell \cos\frac{\theta_\ell}{2} + iA_\ell \sin\frac{\theta_\ell}{2}\sigma_y \right]
\end{equation}

Now, we average over all possible $\theta_\ell\in[0,\pi]$ for sampled Haar-random unitaries under the same set of the error $S_\ell$ and $A_\ell$. For Clements architecture implementing Haar-random unitaries\cite{russell2017direct}, the distribution of splitting angles clusters tightly near $\theta \approx 0$ (the bar state). In this limit, contribution from the antisymmetric $\sin(\theta/2)A_\ell$ is negligible. The error is thus dominated by the symmetric component $||\Delta U_\ell||^2 \approx 2S_\ell^2$.  Since $\alpha_\ell$ and $\beta_\ell$ are statistically independent variables with zero mean and RMS amplitude $\sigma$ ($\mathrm{Var}(\alpha_\ell) = \mathrm{Var}(\beta_\ell) = \sigma^2$), their variances add: $\mathrm{Var}(S_\ell)= \mathrm{Var}(A_\ell) = \mathrm{Var}(\alpha_\ell) + \mathrm{Var}(\beta_\ell) = 2\sigma^2$. Therefore, averaging over all sampled errors, we have $\mathbb{E}[||\Delta U_\ell||^2]=2\mathrm{Var}(S_\ell)=4\sigma^2$.

When $N(N-1)/2$ individual MZIs compose a standard Clements mesh, all the physical errors $\alpha_\ell, \beta_\ell$ can be assumed to be uncorrelated and independently sampled from $\mathcal{N}(0, \sigma)$. The expected error in the first order, $\Delta U_{\mathrm{Clements}}$ scales the Euclidean distance of all independent MZI errors $\{\Delta U_\ell\}$. Thus, the expected squared norm of the error is:

\begin{equation}
\begin{split}
\bar{\mathcal{E}}_{\mathrm{Clements}} & \approx \frac{1}{4N}\mathbb{E}\left[||\Delta U_{\mathrm{Clements}}||^2\right] \\
& \approx \frac{1}{4N}\sum_{\ell=1}^{N(N-1)/2}\mathbb{E}\left[||\Delta U_\ell||^2\right] \\
& = \frac{1}{2}(N-1)\sigma^2 \approx \frac{N}{2}\sigma^2
\end{split}
\end{equation}

\begin{figure*}
    \centering
    \includegraphics[width = \textwidth]{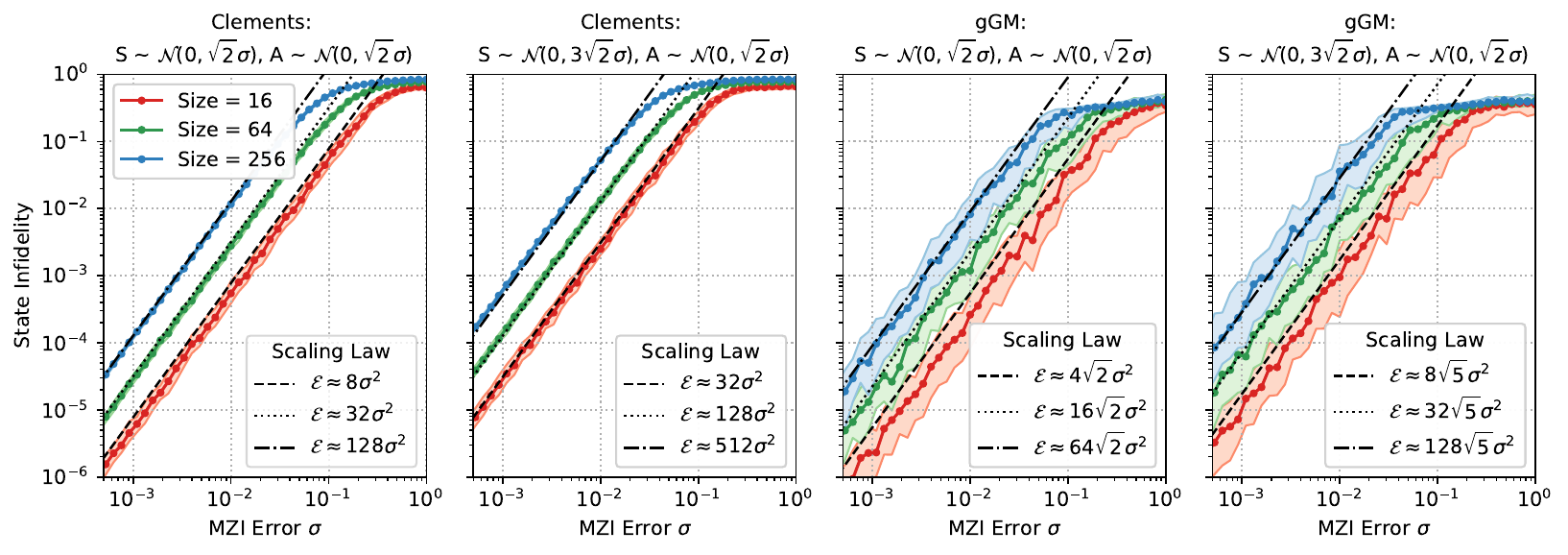}
    \caption{{Scaling of state infidelity as a function of MZI error. Numerical simulations and analytically derived scaling for a single photon state scattered among upto $N = 256$ time bins, with uncorrelated errors (Clements mesh) and correlated errors (gGM). The numerical simulations agree with the analytical solution for large unitaries.}}
    \label{fig:app_scf_scaling}
\end{figure*}

In the recursive time-bin gGM-SCF architecture, the error is defined by a single static vector $\vec{\epsilon} = (S, A)$ fixed for the entire mesh. The total error operator is the coherent sum of propagated local errors:

\begin{equation}
\begin{split}
    \Delta U_{\mathrm{gGM}} \approx & -S \sum_{\ell=1}^{N(N - 1)/2} \underbrace{U_{\mathrm{post}}^{(\ell)} (\cos\tfrac{\theta_\ell}{2}\mathbb{I}^{(\ell)}) U_{\mathrm{pre}}^{(\ell)}}_{K_\ell^{(S)}} \\
    & - iA \sum_{\ell=1}^{N(N - 1)/2} \underbrace{U_{\mathrm{post}}^{(\ell)} (\sin\tfrac{\theta_\ell}{2}\sigma_y^{(\ell)}) U_{\mathrm{pre}}^{(\ell)}}_{K_\ell^{(A)}}
\end{split}
\end{equation}
Unlike the uncorrelated case, the deviation caused by the two error constants $S$ and $A$ accumulates across $N-1$ layers of MZIs. Within each layer, the MZI error applies in parallel. Therefore, the cross-correlation across layers contributes to the overall error. 

From the simulation results shown in Fig.~\ref{fig:app_scf_scaling}, we empirically find that for gGM-SCF, the expected total error scales with the Euclidean norm of the component error terms (quadrature sum) rather than their algebraic sum, which is distinct from the uncorrelated error model used for Clements decomposition.

\begin{equation}
\begin{split}
    \bar{\mathcal{E}}_{\mathrm{gGM}} & \approx \frac{1}{4N}\mathbb{E}\left[||\Delta U_{\mathrm{gGM}}||^2\right] \\
    & \approx \frac{(N-1)}{8} \sqrt{\mathcal{E}_S^2 + \mathcal{E}_A^2} \\
    & \approx \frac{N}{8}\sqrt{\mathrm{Var}(S)^2 + \mathrm{Var}(A)^2}
\end{split}
\end{equation}
For the case where $\mathrm{Var}(S_\ell)= \mathrm{Var}(A_\ell)= 2\sigma^2$, the expected infidelity (error variance) for the generalized Green Machine scales as:
\begin{equation}
\bar{\mathcal{E}}_{\mathrm{gGM}}\approx \frac{N}{2\sqrt{2}}\sigma^2
\end{equation}
which gives us the $\sqrt{2}$ improvement in the mean fidelity seen in Fig. 2(b) of the main text. Furthermore, a broader distribution of infidelity (larger variances) is also seen in the histograms from Fig. 2(b) and Fig.~\ref{fig:app_scf_scaling}. 

Finally, we extend this single-particle analysis to the case of an input state $\ket{\psi}$ containing $N_{\mathrm{ph}}$ photons. In a linear lossless optical network, the transformation is particle-conserving; a coherent phase error $\phi$ in the mesh creates a phase rotation $e^{i\phi}$ on each traversing particle. For the many-body state, this accumulates into a global phase shift $U(\phi) \sim e^{i N_{\mathrm{ph}} \phi}$. The total error operator $\Delta U$ is determined by the derivative of this transformation with respect to the phase error, $\frac{\partial U}{\partial \phi}$. In the second-quantized formalism, this promotes the error generator to the total photon number operator $\hat{n} = \sum \hat{a}^\dagger_i \hat{a}_i$. Since the infidelity is dominated by the variance of this error generator, and errors accumulate additively across the independent excitations in the limit of small perturbations, the total infidelity scales linearly with the photon number for both correlated and uncorrelated error models..  Combining this particle scaling with the mesh scaling derived above yields the total infidelity:
\begin{equation}
    \mathcal{E} \propto N_{\mathrm{ph}}(N-1)\sigma^{2} \approx N_{\mathrm{ph}}N\sigma^{2}
\end{equation}

\section*{Appendix C: Inaccessible states from intra-MZI imbalance of loss\label{sec:appendix_3}}

\begin{figure}
    \centering
    \includegraphics[width =1.1\columnwidth]{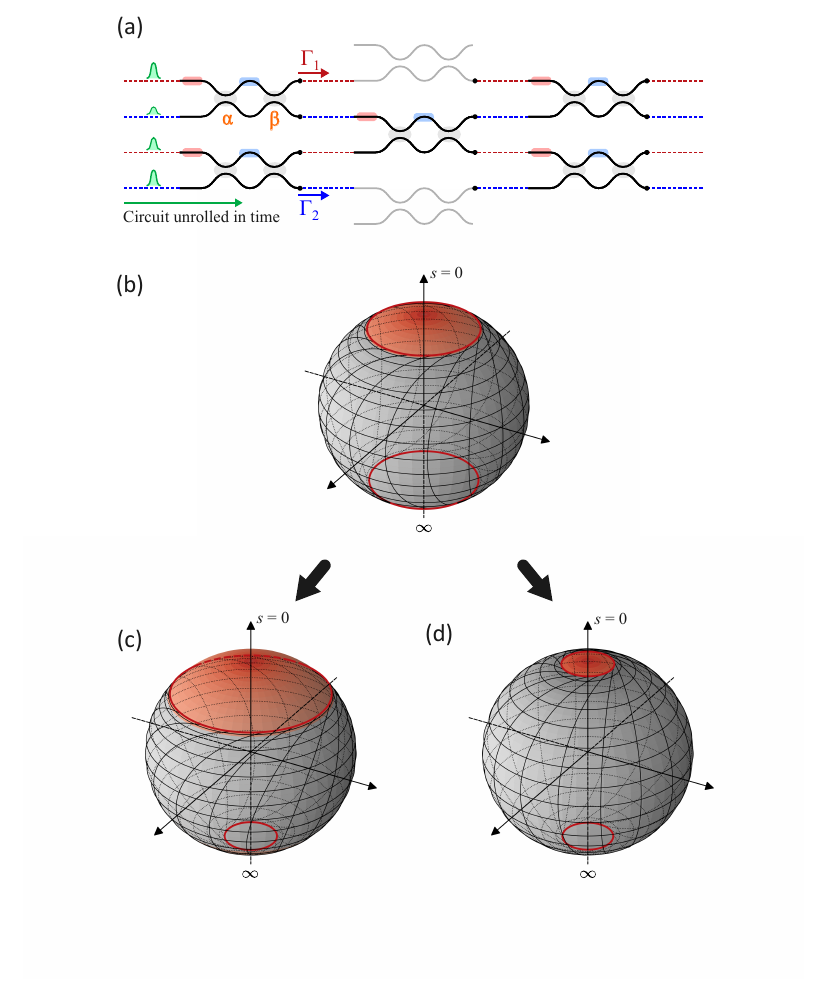}    \caption{An illustration of accessible regions of tunable beamsplitter's splitting ratios restricted by loss and coherent errors. (a) Clement mesh circuit implemented by the recursive structure of the generalized Green Machine unrolled in time, identifying the two imbalanced loss factors $\Gamma_1$ and $\Gamma_2$ taken into account. (b) Riemann sphere representation of accessible splitting ratios, with the forbidden regions shown in red. Purely unbalanced loss (transmission coefficients chosen to be $(\Gamma_{1} = 0.5, \Gamma_{2} = 0.9)$ for illustrative purposes) introduces forbidden regions around the poles. (c) Coherent errors (chosen to be  $(\alpha = 0.15, \beta = 0.0)$) perturb the forbidden regions around the poles asymmetrically, which could happen in spatially meshed linear optical circuits. (d) Lower degree of imbalance in the loss [transmission coefficients chosen to be $(\Gamma_{1} = 0.7, \Gamma_{2} = 0.9)$] on each arm shrinks the forbidden regions.}
    \label{fig:GM_Riemann}
\end{figure}

In practice, the unbalanced transmission coefficients on each arm of the MZI, denoted by $(\Gamma_{1}, \Gamma_{2})$ in Fig.~\ref{fig:GM_Riemann}(a), resulting in a reduced fraction of the unitary group that can be realized. This drop in expressivity results in groups of states that cannot be prepared by the linear optical circuit. These inaccessible output states that arise as a result of coherent errors and unbalanced loss can be determined by evaluating the range of admissible splitting ratios implemented by the end-to-end transfer matrix of our single MZI beamsplitter~\cite{hamerly2022asymptotically}. For a given $2 \times 2$ transfer matrix $U$ implemented on a set of time-bin modes, the complex-valued splitting ratio is defined as $s = U_{11}/U_{12}$. The range of admissible splitting ratios in the presence of splitter error $(\alpha, \beta)$ and unbalanced transmission coefficients $(\Gamma_{1}, \Gamma_{2})$ on each arm reduces to

\begin{widetext}
\begin{multline}
    \frac{\Gamma_{1} \left[ \mathrm{cos}|\alpha - \beta| - \mathrm{sin}|\alpha + \beta| \right] - \Gamma_{2} \left[ \mathrm{cos}|\alpha - \beta| + \mathrm{sin}|\alpha + \beta| \right]}{\Gamma_{1} \left[ \mathrm{cos}|\alpha + \beta| - \mathrm{sin}|\alpha - \beta| \right] + \Gamma_{2} \left[ \mathrm{cos}|\alpha + \beta| + \mathrm{sin}|\alpha - \beta| \right]} \leq |s| \\
    \leq ~ \frac{\Gamma_{1} \left[ \mathrm{cos}|\alpha - \beta| - \mathrm{sin}|\alpha + \beta| \right] + \Gamma_{2} \left[ \mathrm{cos}|\alpha - \beta| + \mathrm{sin}|\alpha + \beta| \right]}{\Gamma_{1} \left[ \mathrm{cos}|\alpha + \beta| - \mathrm{sin}|\alpha - \beta| \right] - \Gamma_{2} \left[ \mathrm{cos}|\alpha + \beta| + \mathrm{sin}|\alpha - \beta| \right]}
    \label{eqn:splitting}
\end{multline}
\end{widetext}
which reduces to $\mathrm{tan}|\alpha + \beta| \leq |s| \leq \mathrm{cot}|\alpha - \beta|$ in the lossless setting~\cite{hamerly2022asymptotically} and $\frac{\Gamma_{1} - \Gamma_{2}}{\Gamma_{1} + \Gamma_{2}} \leq |s| \leq \frac{\Gamma_{1} + \Gamma_{2}}{\Gamma_{1} - \Gamma_{2}}$ in the absence of coherent errors.

Accessible splitting ratios \textit{over a single stage} in the presence of these errors can be visualized on a Riemann Sphere, illustrated in Fig.~\ref{fig:GM_Riemann}. Under purely unbalanced loss where ($\Gamma_{1} = 0.5, \Gamma_{2} = 0.9$ chosen for illustrative purposes), forbidden regions near the poles emerge, indicated in red in Fig.~\ref{fig:GM_Riemann}(b), showing regions where perfect nulling of the power is impossible. In the presence of both unbalanced loss and coherent errors, the forbidden regions that emerge are asymmetric, given by Eq.~\eqref{eqn:splitting}. Fig.~\ref{fig:GM_Riemann}(c) illustrates this case with $(\alpha = 0.15, \beta = 0.0)$, where the inaccessible region at the $s = 0$ pole expands, while the region at the $s = \infty$ pole contracts. This suggests that the forbidden region at one of the poles can be eliminated entirely, allowing the realization of the perfect cross state for improved matrix fidelity.~\cite{miller2015perfect, suzuki2015ultra, hamerly2022asymptotically}.

When the degree of imbalance in loss on each arm is reduced (i.e., as $\Gamma_{1}/\Gamma_{2} \to 1$), the inaccessible regions shrink, shown in Fig.~\ref{fig:GM_Riemann}(d). Perfectly balanced loss eliminates the forbidden regions completely, allowing access to the full range of splitting ratios while resulting in only a scaling down of the output power. Imbalanced loss arising from interfacing each time bin with delay lines asymmetrically \textit{over multiple stages} (such as the Clements configuration shown in Fig.~\ref{fig:GM_Riemann}(a)) cascades into a complex chain of errors, further reducing circuit expressivity. This expressivity can be parameterized by two coherent-error parameters $\alpha$ and $\beta$, two intra-MZI loss parameters $\Gamma_1$ and $\Gamma_2$, and two delay line loss rates, which is the subject of our future work.

\section*{Appendix D: Large-Scale Multiplexing for Tensor Operations\label{sec:appendix_4}}

\begin{figure}
    \centering
    \includegraphics[width = \columnwidth]{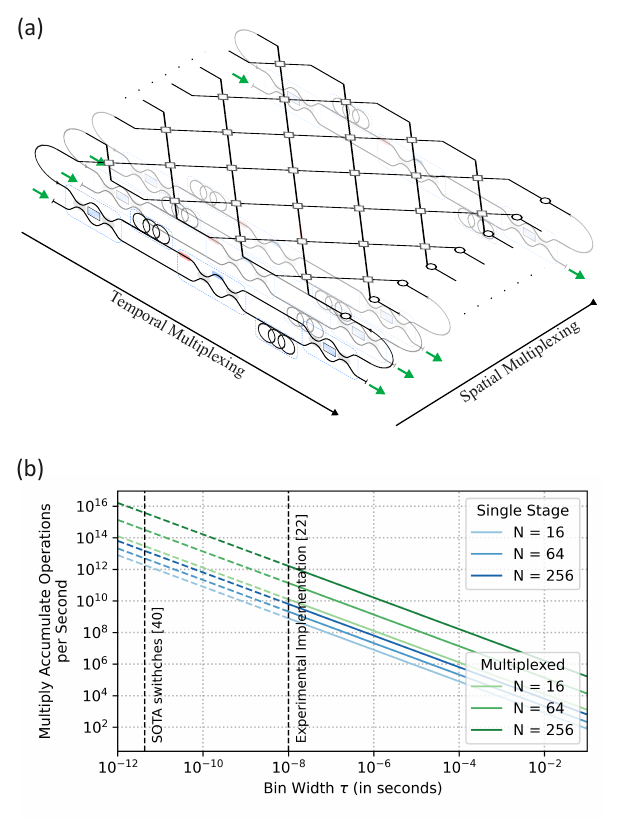} \caption{Multiplexed architecture and computational speed for high-dimensional tensor operations. (a)  Multiplexed architecture by cascading a single-stage generalized Green Machine for high-dimensional unitary transformations for data encoded in spatial and temporal degrees of freedom. (b) Number of multiply accumulate operations per second in matrix-vector multiplication as a function of the bin width, as decided by the switching speed. The experimental implementation of the Green Machine is marked in ref.~\cite{cui2025superadditive} at a bin width of $10~\mathrm{ns}$. State-of-the-art switching speeds from ref.~\cite{bouchard2024programmable} is marked, at bin widths of $\approx 4.3~\mathrm{ps}$.}
    \label{fig:app_speedometer}
\end{figure}

While the results in this manuscript have mainly focused on applications involving quantum computing, this architecture is also well suited to implement the large-scale and high-dimensional unitaries necessary for machine learning accelerators. A metric commonly used to evaluate the performance of these accelerators is the number of multiply accumulate operations per second (MACs). Here, we evaluate the computational speed of matrix-vector multiplications implemented by the generalized Green Machine as a function of the width of the time bins used. 

 Fig.~\ref{fig:app_speedometer}(a) illustrates a schematic for a spatially and temporally multiplexed Green Machine architecture, consisting of multiple stages of the generalized Green Machine introduced in the main text of this manuscript. By stacking $M$ individual stages together, where each stage operated on $N$ time-bin modes, this processor can be used to perform $M \times M \times N$ tensor operations. Fig.~\ref{fig:app_speedometer}(b) plots the computational speed of a single stage and the multiplexed stages as a function of the width of the time bins. We mark the speeds of each architecture using the dotted line at a previous experimental implementation~\cite{cui2025superadditive} from the authors and state-of-the-art switching speeds~\cite{bouchard2024programmable}.

\section*{Appendix E: Performance Under Hardware Error Correction\label{sec:appendix_5}}

\begin{figure}
    \centering
    \includegraphics[width = \columnwidth]{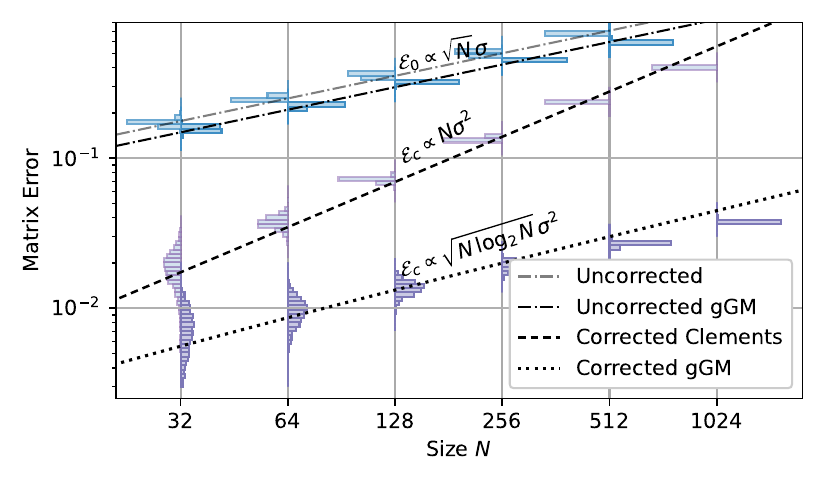} \caption{Scaling of the matrix errors for the Clements architecture (with uncorrelated errors $\sigma=0.02$), the generalized Green Machine (with correlated errors $\sigma=0.02$) in the absence and presence of hardware error correction. The improvement from correlated errors are shown in the blue histograms, where the Green Machine provides a constant improvement over the Clements mesh. Additional use of hardware error-correction techniques provides a further significant scaling improvement, as indicated in purple.}
    \label{fig:app_corrected_scaling}
\end{figure}

The main text of this manuscript demonstrates an improvement in the robustness to coherent errors of the generalized Green Machine compared to the spatial mode Clements architecture. This improvement arises from the correlated nature of errors that is a result of the repeated use of a single MZI. We show in figure 2 of the main text that under correlated errors, the scaling law for the state fidelity/matrix error remains identical, with an improvement in the constant prefactor. In Fig.~\ref{fig:app_corrected_scaling} we plot the scaling laws to compare the error tolerance of the Clements architecture to the generalized Green Machine under hardware error correction~\cite{bandyopadhyay2021hardware}. In the figure, the darker distributions on the right plot the matrix error for the Green Machine, while the lighter distributions to the left plot the matrix error for the Clements mesh. The blue histograms illustrate the improvement arising from the correlated nature of errors. Additional use of error correction techniques provides a significant scaling improvement, as indicated by the purple histograms.

\clearpage

\bibliography{arXiv}

@article{liu2025quantum,
  title={Quantum learning advantage on a scalable photonic platform},
  author={Liu, Zheng-Hao and Brunel, Romain and {\O}stergaard, Emil EB and Cordero, Oscar and Chen, Senrui and Wong, Yat and Nielsen, Jens AH and Bregnsbo, Axel B and Zhou, Sisi and Huang, Hsin-Yuan and others},
  journal={Science},
  volume={389},
  number={6767},
  pages={1332--1335},
  year={2025},
  doi={10.1126/science.adv2560},
  publisher={American Association for the Advancement of Science}
}

@article{he2019low,
  title={Low-loss fiber-to-chip interface for lithium niobate photonic integrated circuits},
  author={He, Lingyan and Zhang, Mian and Shams-Ansari, Amirhassan and Zhu, Rongrong and Wang, Cheng and Marko, Lon{\v{c}}ar},
  journal={Optics Letters},
  volume={44},
  number={9},
  pages={2314--2317},
  year={2019},
  doi = {10.1364/OL.44.002314},
  publisher={Optica Publishing Group}
}

@inproceedings{arnold2023free,
  title={Free-space photonic quantum memory},
  author={Arnold, Nathan T and Victora, Michelle and Goggin, Michael E and Kwiat, Paul G},
  booktitle={Quantum Computing, Communication, and Simulation III},
  volume={12446},
  pages={25--30},
  year={2023},
  city = {San Francisco, California, United States},
  doi = {10.1117/12.2649350},
  organization={SPIE}
}

@article{chang2017heterogeneous,
  title={Heterogeneous integration of lithium niobate and silicon nitride waveguides for wafer-scale photonic integrated circuits on silicon},
  author={Chang, Lin and Pfeiffer, Martin HP and Volet, Nicolas and Zervas, Michael and Peters, Jon D and Manganelli, Costanza L and Stanton, Eric J and Li, Yifei and Kippenberg, Tobias J and Bowers, John E},
  journal={Optics Letters},
  volume={42},
  number={4},
  pages={803--806},
  year={2017},
  doi = {10.1364/OL.42.000803},
  publisher={Optica Publishing Group}
}

@article{psiquantum2025manufacturable,
  title={A manufacturable platform for photonic quantum computing},
  journal={Nature},
  pages={1--3},
  year={2025},
  doi = {10.1038/s41586-025-08820-7},
  publisher={Nature Publishing Group UK London}
}

@article{pant2019percolation,
  title={Percolation thresholds for photonic quantum computing},
  author={Pant, Mihir and Towsley, Don and Englund, Dirk and Guha, Saikat},
  journal={Nature Communications},
  volume={10},
  number={1},
  pages={1070},
  year={2019},
  doi = {10.1038/s41467-019-08948-x},
  publisher={Nature Publishing Group UK London}
}

@article{hauser2025boosted,
  title={Boosted Bell-state measurements for photonic quantum computation},
  author={Hauser, Nico and Bayerbach, Matthias J and D’Aurelio, Simone E and Weber, Raphael and Santandrea, Matteo and Kumar, Shreya P and Dhand, Ish and Barz, Stefanie},
  journal={npj Quantum Information},
  volume={11},
  number={1},
  pages={41},
  year={2025},
  doi = {10.1038/s41534-025-00986-2},
  publisher={Nature Publishing Group UK London}
}

@article{dhara2023entangling,
  title={Entangling quantum memories via heralded photonic Bell measurement},
  author={Dhara, Prajit and Englund, Dirk and Guha, Saikat},
  journal={Physical Review Research},
  volume={5},
  number={3},
  pages={033149},
  year={2023},
  doi = {10.1103/PhysRevResearch.5.033149},
  publisher={APS}
}

@article{ewert20143,
  title={3/4-efficient bell measurement with passive linear optics and unentangled ancillae},
  author={Ewert, Fabian and van Loock, Peter},
  journal={Physical Review Letters},
  volume={113},
  number={14},
  pages={140403},
  year={2014},
  doi = {10.1103/PhysRevLett.113.140403},
  publisher={APS}
}

@article{anderson1958absence,
  title={Absence of diffusion in certain random lattices},
  author={Anderson, Philip W},
  journal={Physical Review},
  volume={109},
  number={5},
  pages={1492},
  year={1958},
  doi = {10.1103/PhysRev.109.1492},
  publisher={APS}
}

@article{madsen2022quantum,
  title={Quantum computational advantage with a programmable photonic processor},
  author={Madsen, Lars S and Laudenbach, Fabian and Askarani, Mohsen Falamarzi and Rortais, Fabien and Vincent, Trevor and Bulmer, Jacob FF and Miatto, Filippo M and Neuhaus, Leonhard and Helt, Lukas G and Collins, Matthew J and others},
  journal={Nature},
  volume={606},
  number={7912},
  pages={75--81},
  year={2022},
  doi = {10.1038/s41586-022-04725-x},
  publisher={Nature Publishing Group UK London}
}

@article{zhong2020quantum,
  title={Quantum computational advantage using photons},
  author={Zhong, Han-Sen and Wang, Hui and Deng, Yu-Hao and Chen, Ming-Cheng and Peng, Li-Chao and Luo, Yi-Han and Qin, Jian and Wu, Dian and Ding, Xing and Hu, Yi and others},
  journal={Science},
  volume={370},
  number={6523},
  pages={1460--1463},
  year={2020},
  doi = {10.1126/science.abe8770},
  publisher={American Association for the Advancement of Science}
}

@article{xia2021quantum,
  title={Quantum-enhanced data classification with a variational entangled sensor network},
  author={Xia, Yi and Li, Wei and Zhuang, Quntao and Zhang, Zheshen},
  journal={Physical Review X},
  volume={11},
  number={2},
  pages={021047},
  year={2021},
  doi = {10.1103/PhysRevX.11.021047},
  publisher={APS}
}

@article{luo2019quantum,
  title={Quantum teleportation in high dimensions},
  author={Luo, Yi-Han and Zhong, Han-Sen and Erhard, Manuel and Wang, Xi-Lin and Peng, Li-Chao and Krenn, Mario and Jiang, Xiao and Li, Li and Liu, Nai-Le and Lu, Chao-Yang and others},
  journal={Physical Review Letters},
  volume={123},
  number={7},
  pages={070505},
  year={2019},
  doi = {10.1103/PhysRevLett.123.070505},
  publisher={APS}
}

@article{cui2020high,
  title={High-dimensional frequency-encoded quantum information processing with passive photonics and time-resolving detection},
  author={Cui, Chaohan and Seshadreesan, Kaushik P and Guha, Saikat and Fan, Linran},
  journal={Physical Review Letters},
  volume={124},
  number={19},
  pages={190502},
  year={2020},
  doi = {10.1103/PhysRevLett.124.190502},
  publisher={APS}
}

@article{zhang2019quantum,
  title={Quantum teleportation of photonic qudits using linear optics},
  author={Zhang, Chenyu and Chen, JF and Cui, Chaohan and Dowling, Jonathan P and Ou, ZY and Byrnes, Tim},
  journal={Physical Review A},
  volume={100},
  number={3},
  pages={032330},
  year={2019},
  doi = {10.1103/PhysRevA.100.032330},
  publisher={APS}
}

@article{guha2011structured,
  title={Structured optical receivers to attain superadditive capacity and the Holevo limit},
  author={Guha, Saikat},
  journal={Physical Review Letters},
  volume={106},
  number={24},
  pages={240502},
  year={2011},
  doi = {10.1103/PhysRevLett.106.240502},
  publisher={APS}
}

@article{calsamiglia2001maximum,
  title={Maximum efficiency of a linear-optical Bell-state analyzer},
  author={Calsamiglia, John and L{\"u}tkenhaus, Norbert},
  journal={Applied Physics B},
  volume={72},
  pages={67--71},
  year={2001},
  doi = {10.1007/s003400000484},
  publisher={Springer}
}

@article{melkozerov2024analysis,
  title={Analysis of optical loss thresholds in the fusion-based quantum computing architecture},
  author={Melkozerov, Aleksandr and Avanesov, Ashot and Dyakonov, Ivan and Straupe, Stanislav},
  journal={APL Quantum},
  volume={1},
  number={3},
  year={2024},
  doi = {10.1063/5.0214728},
  publisher={AIP Publishing}
}

@article{rosati2024joint,
  title={Joint-detection learning for optical communication at the quantum limit},
  author={Rosati, Matteo and Solana, Albert},
  journal={Optica Quantum},
  volume={2},
  number={6},
  pages={390--396},
  year={2024},
  doi = {10.1364/OPTICAQ.521637},
  publisher={Optica Publishing Group}
}

@article{rosati2025fourier,
  title={A Fourier machine for quantum optical communications},
  author={Rosati, Matteo and Cincotti, Gabriella},
  journal={Journal of Lightwave Technology},
  volume={43},
  number={8},
  pages={3770--3776},
  doi = {10.1109/JLT.2025.3530535},
  year={2025}
}

@article{clements2016optimal,
  title={{Optimal design for universal multiport interferometers}},
  author={Clements, William R and Humphreys, Peter C and Metcalf, Benjamin J and Kolthammer, W Steven and Walmsley, Ian A},
  journal={Optica},
  volume={3},
  number={12},
  pages={1460--1465},
  year={2016},
  doi = {10.1364/OPTICA.3.001460},
  publisher={Optical Society of America}
}

@article{reck1994experimental,
  title={{Experimental realization of any discrete unitary operator}},
  author={Reck, Michael and Zeilinger, Anton and Bernstein, Herbert J and Bertani, Philip},
  journal={Physical Review Letters},
  volume={73},
  number={1},
  pages={58},
  year={1994},
  doi = {10.1103/PhysRevLett.73.58},
  publisher={APS}
}

@article{basani2023self,
  title={A self-similar sine--cosine fractal architecture for multiport interferometers},
  author={Basani, Jasvith Raj and Vadlamani, Sri Krishna and Bandyopadhyay, Saumil and Englund, Dirk R and Hamerly, Ryan},
  journal={Nanophotonics},
  volume={12},
  number={5},
  pages={975--984},
  year={2023},
  doi = {10.1515/nanoph-2022-0525},
  publisher={De Gruyter}
}

@article{hamerly2022asymptotically,
  title={Asymptotically fault-tolerant programmable photonics},
  author={Hamerly, Ryan and Bandyopadhyay, Saumil and Englund, Dirk},
  journal={Nature Communications},
  volume={13},
  number={1},
  pages={6831},
  year={2022},
  doi = {10.1038/s41467-022-34308-3},
  publisher={Nature Publishing Group UK London}
}

@article{hamerly2022stability,
  title={{Stability of self-configuring large multiport interferometers}},
  author={Hamerly, Ryan and Bandyopadhyay, Saumil and Englund, Dirk},
  journal={Physical Review Applied},
  volume={18},
  number={2},
  pages={024018},
  year={2022},
  doi = {10.1103/PhysRevApplied.18.024018},
  publisher={APS}
}

@article{hamerly2022accurate,
  title={{Accurate self-configuration of rectangular multiport interferometers}},
  author={Hamerly, Ryan and Bandyopadhyay, Saumil and Englund, Dirk},
  journal={Physical Review Applied},
  volume={18},
  number={2},
  pages={024019},
  year={2022},
  doi = {10.1103/PhysRevApplied.18.024019},
  publisher={APS}
}

@article{bandyopadhyay2021hardware,
  title={{Hardware error correction for programmable photonics}},
  author={Bandyopadhyay, Saumil and Hamerly, Ryan and Englund, Dirk},
  journal={Optica},
  volume={8},
  number={10},
  pages={1247--1255},
  year={2021},
  doi = {10.1364/OPTICA.424052},
  publisher={Optical Society of America}
}

@article{carolan2015universal,
  title={{Universal linear optics}},
  author={Carolan, Jacques and Harrold, Christopher and Sparrow, Chris and Mart{\'\i}n-L{\'o}pez, Enrique and Russell, Nicholas J and Silverstone, Joshua W and Shadbolt, Peter J and Matsuda, Nobuyuki and Oguma, Manabu and Itoh, Mikitaka and others},
  journal={Science},
  volume={349},
  number={6249},
  pages={711--716},
  year={2015},
  doi={10.1126/science.aab3642},
  publisher={American Association for the Advancement of Science}
}

@article{shen2017deep,
  title={{Deep learning with coherent nanophotonic circuits}},
  author={Shen, Yichen and Harris, Nicholas C and Skirlo, Scott and Prabhu, Mihika and Baehr-Jones, Tom and Hochberg, Michael and Sun, Xin and Zhao, Shijie and Larochelle, Hugo and Englund, Dirk and others},
  journal={Nature Photonics},
  volume={11},
  number={7},
  pages={441--446},
  year={2017},
  doi = {10.1038/nphoton.2017.93},
  publisher={Nature Publishing Group}
}

@article{hamerly2019large,
  title={{Large-scale optical neural networks based on photoelectric multiplication}},
  author={Hamerly, Ryan and Bernstein, Liane and Sludds, Alexander and Solja{\v{c}}i{\'c}, Marin and Englund, Dirk},
  journal={Physical Review X},
  volume={9},
  number={2},
  pages={021032},
  year={2019},
  doi={10.1103/PhysRevX.9.021032},
  publisher={APS}
}

@article{carolan2020variational,
  title={Variational quantum unsampling on a quantum photonic processor},
  author={Carolan, Jacques and Mohseni, Masoud and Olson, Jonathan P and Prabhu, Mihika and Chen, Changchen and Bunandar, Darius and Niu, Murphy Yuezhen and Harris, Nicholas C and Wong, Franco NC and Hochberg, Michael and others},
  journal={Nature Physics},
  volume={16},
  number={3},
  pages={322--327},
  year={2020},
  doi = {10.1038/s41567-019-0747-6},
  publisher={Nature Publishing Group UK London}
}

@article{harris2017quantum,
  title={{Quantum transport simulations in a programmable nanophotonic processor}},
  author={Harris, Nicholas C and Steinbrecher, Gregory R and Prabhu, Mihika and Lahini, Yoav and Mower, Jacob and Bunandar, Darius and Chen, Changchen and Wong, Franco NC and Baehr-Jones, Tom and Hochberg, Michael and others},
  journal={Nature Photonics},
  volume={11},
  number={7},
  pages={447--452},
  year={2017},
  doi={10.1038/nphoton.2017.95},
  publisher={Nature Publishing Group}
}

@article{qiang2018large,
  title={{Large-scale silicon quantum photonics implementing arbitrary two-qubit processing}},
  author={Qiang, Xiaogang and Zhou, Xiaoqi and Wang, Jianwei and Wilkes, Callum M and Loke, Thomas and O’Gara, Sean and Kling, Laurent and Marshall, Graham D and Santagati, Raffaele and Ralph, Timothy C and others},
  journal={Nature Photonics},
  volume={12},
  number={9},
  pages={534--539},
  year={2018},
  doi={10.1038/s41566-018-0236-y},
  publisher={Nature Publishing Group}
}

@article{sparrow2018simulating,
  title={{Simulating the vibrational quantum dynamics of molecules using photonics}},
  author={Sparrow, Chris and Mart{\'\i}n-L{\'o}pez, Enrique and Maraviglia, Nicola and Neville, Alex and Harrold, Christopher and Carolan, Jacques and Joglekar, Yogesh N and Hashimoto, Toshikazu and Matsuda, Nobuyuki and O’Brien, Jeremy L and others},
  journal={Nature},
  volume={557},
  number={7707},
  pages={660--667},
  year={2018},
  doi={10.1038/s41586-018-0152-9},
  publisher={Nature Publishing Group}
}

@article{basani2024all,
  title={All-photonic artificial-neural-network processor via nonlinear optics},
  author={Basani, Jasvith Raj and Heuck, Mikkel and Englund, Dirk R and Krastanov, Stefan},
  journal={Physical Review Applied},
  volume={22},
  number={1},
  pages={014009},
  year={2024},
  doi = {10.1103/PhysRevApplied.22.014009},
  publisher={APS}
}

@article{bandyopadhyay2024single,
  title={Single-chip photonic deep neural network with forward-only training},
  author={Bandyopadhyay, Saumil and Sludds, Alexander and Krastanov, Stefan and Hamerly, Ryan and Harris, Nicholas and Bunandar, Darius and Streshinsky, Matthew and Hochberg, Michael and Englund, Dirk},
  journal={Nature Photonics},
  volume={18},
  number={12},
  pages={1335--1343},
  year={2024},
  doi = {10.1038/s41566-024-01567-z},
  publisher={Nature Publishing Group}
}

@article{bao2023very,
  title={Very-large-scale integrated quantum graph photonics},
  author={Bao, Jueming and Fu, Zhaorong and Pramanik, Tanumoy and Mao, Jun and Chi, Yulin and Cao, Yingkang and Zhai, Chonghao and Mao, Yifei and Dai, Tianxiang and Chen, Xiaojiong and others},
  journal={Nature Photonics},
  pages={1--9},
  year={2023},
  doi={10.1038/s41566-023-01187-z},
  publisher={Nature Publishing Group UK London}
}

@article{ewaniuk2023imperfect,
  title={Imperfect Quantum Photonic Neural Networks},
  author={Ewaniuk, Jacob and Carolan, Jacques and Shastri, Bhavin J and Rotenberg, Nir},
  journal={Advanced Quantum Technologies},
  volume={6},
  number={3},
  pages={2200125},
  year={2023},
  doi={10.1002/qute.202200125},
  publisher={Wiley Online Library}
}

@article{vadlamani2023transferable,
  title={Transferable learning on analog hardware},
  author={Vadlamani, Sri Krishna and Englund, Dirk and Hamerly, Ryan},
  journal={Science Advances},
  volume={9},
  number={28},
  pages={eadh3436},
  year={2023},
  doi = {10.1126/sciadv.adh3436},
  publisher={American Association for the Advancement of Science}
}

@article{burgwal2017using,
  title={{Using an imperfect photonic network to implement random unitaries}},
  author={Burgwal, Roel and Clements, William R and Smith, Devin H and Gates, James C and Kolthammer, W Steven and Renema, Jelmer J and Walmsley, Ian A},
  journal={Optics Express},
  volume={25},
  number={23},
  pages={28236--28245},
  year={2017},
  doi={10.1364/OE.25.028236},
  publisher={Optica Publishing Group}
}

@article{kumar2021mitigating,
  title={{Mitigating linear optics imperfections via port allocation and compilation}},
  author={Kumar, Shreya P and Neuhaus, Leonhard and Helt, Lukas G and Qi, Haoyu and Morrison, Blair and Mahler, Dylan H and Dhand, Ish},
  journal={arXiv preprint},
  doi={arXiv:2103.03183},
  year={2021}
}

@article{romero2017quantum,
  title={Quantum autoencoders for efficient compression of quantum data},
  author={Romero, Jonathan and Olson, Jonathan P and Aspuru-Guzik, Alan},
  journal={Quantum Science and Technology},
  volume={2},
  number={4},
  pages={045001},
  year={2017},
  doi = {10.1088/2058-9565/aa8072},
  publisher={IOP Publishing}
}

@article{chi2022programmable,
  title={A programmable qudit-based quantum processor},
  author={Chi, Yulin and Huang, Jieshan and Zhang, Zhanchuan and Mao, Jun and Zhou, Zinan and Chen, Xiaojiong and Zhai, Chonghao and Bao, Jueming and Dai, Tianxiang and Yuan, Huihong and others},
  journal={Nature Communications},
  volume={13},
  number={1},
  pages={1166},
  year={2022},
  doi = {10.1038/s41467-022-28767-x},
  publisher={Nature Publishing Group UK London}
}

@article{hamerly2025toward,
  title={Toward the information-theoretic limit of programmable photonics},
  author={Hamerly, Ryan and Basani, Jasvith R and Sludds, Alexander and Vadlamani, Sri K and Englund, Dirk},
  journal={APL Photonics},
  volume={10},
  number={11},
  year={2025},
  doi={10.1063/5.0269741},
  publisher={AIP Publishing}
}

@inproceedings{hamerly2023multiplexing,
  title={Multiplexing methods for scaling up photonic logic},
  author={Hamerly, Ryan and Bandyopadhyay, Saumil and Sludds, Alexander and Chen, Zaijun and Bernstein, Liane and Vadlamani, Sri Krishna and Basani, Jasvith and Davis, Ronald and Englund, Dirk},
  booktitle={AI and Optical Data Sciences IV},
  volume={12438},
  pages={1243802},
  year={2023},
  city = {San Francisco, California, United States},
  doi = {10.1117/12.2650902},
  organization={SPIE}
}

@article{yu2023heavy,
  title={Heavy tails and pruning in programmable photonic circuits for universal unitaries},
  author={Yu, Sunkyu and Park, Namkyoo},
  journal={Nature Communications},
  volume={14},
  number={1},
  pages={1853},
  year={2023},
  doi = {10.1038/s41467-023-37611-9},
  publisher={Nature Publishing Group UK London}
}

@article{cui2025superadditive,
  title={Superadditive communication with the green machine as a practical demonstration of nonlocality without entanglement},
  author={Cui, Chaohan and Postlewaite, Jack and Saif, Babak N and Fan, Linran and Guha, Saikat},
  journal={Nature Communications},
  volume={16},
  number={1},
  pages={3760},
  year={2025},
  doi = {10.1038/s41467-025-59107-4},
  publisher={Nature Publishing Group UK London}
}

@article{miller2015perfect,
  title={Perfect optics with imperfect components},
  author={Miller, David AB},
  journal={Optica},
  volume={2},
  number={8},
  pages={747--750},
  year={2015},
  doi = {10.1364/OPTICA.2.000747},
  publisher={Optical Society of America}
}

@article{suzuki2015ultra,
  title={Ultra-high-extinction-ratio 2$\times$ 2 silicon optical switch with variable splitter},
  author={Suzuki, Keijiro and Cong, Guangwei and Tanizawa, Ken and Kim, Sang-Hun and Ikeda, Kazuhiro and Namiki, Shu and Kawashima, Hitoshi},
  journal={Optics Express},
  volume={23},
  number={7},
  pages={9086--9092},
  year={2015},
  doi = {10.1364/OE.23.009086},
  publisher={Optical Society of America}
}

@article{basani2025universal,
  title={Universal logical quantum photonic neural network processor via cavity-assisted interactions},
  author={Basani, Jasvith Raj and Niu, Murphy Yuezhen and Waks, Edo},
  journal={npj Quantum Information},
  volume={11},
  number={1},
  pages={142},
  year={2025},
  doi={10.1038/s41534-025-01096-9},
  publisher={Nature Publishing Group UK London}
}

@article{ahmed2025universal,
  title={Universal photonic artificial intelligence acceleration},
  author={Ahmed, Sufi R and Baghdadi, Reza and Bernadskiy, Mikhail and Bowman, Nate and Braid, Ryan and Carr, Jim and Chen, Chen and Ciccarella, Pietro and Cole, Matthew and Cooke, John and others},
  journal={Nature},
  volume={640},
  number={8058},
  pages={368--374},
  year={2025},
  doi = {10.1038/s41586-025-08854-x},
  publisher={Nature Publishing Group UK London}
}

@article{hua2025integrated,
  title={An integrated large-scale photonic accelerator with ultralow latency},
  author={Hua, Shiyue and Divita, Erwan and Yu, Shanshan and Peng, Bo and Roques-Carmes, Charles and Su, Zhan and Chen, Zhang and Bai, Yanfei and Zou, Jinghui and Zhu, Yunpeng and others},
  journal={Nature},
  volume={640},
  number={8058},
  pages={361--367},
  year={2025},
  doi = {10.1038/s41586-025-08786-6},
  publisher={Nature Publishing Group UK London}
}

@article{ou2025hypermultiplexed,
  title={Hypermultiplexed integrated photonics--based optical tensor processor},
  author={Ou, Shaoyuan and Xue, Kaiwen and Zhou, Lian and Lee, Chun-ho and Sludds, Alexander and Hamerly, Ryan and Zhang, Ke and Feng, Hanke and Yu, Yue and Kopparapu, Reshma and others},
  journal={Science Advances},
  volume={11},
  number={23},
  pages={eadu0228},
  year={2025},
  doi={10.1126/sciadv.adu0228},
  publisher={American Association for the Advancement of Science}
}

@article{mosca2002novel,
  title={A novel design method for Blass matrix beam-forming networks},
  author={Mosca, Stefano and Bilotti, Filiberto and Toscano, Alessandro and Vegni, Lucio},
  journal={IEEE Transactions on Antennas and Propagation},
  volume={50},
  number={2},
  pages={225--232},
  year={2002},
  doi = {10.1109/8.997999},
  publisher={IEEE}
}

@article{taballione20198,
  title={8$\times$ 8 reconfigurable quantum photonic processor based on silicon nitride waveguides},
  author={Taballione, Caterina and Wolterink, Tom AW and Lugani, Jasleen and Eckstein, Andreas and Bell, Bryn A and Grootjans, Robert and Visscher, Ilka and Geskus, Dimitri and Roeloffzen, Chris GH and Renema, Jelmer J and others},
  journal={Optics Express},
  volume={27},
  number={19},
  pages={26842--26857},
  year={2019},
  doi = {10.1364/OE.27.026842},
  publisher={Optical Society of America}
}

@article{suzuki2014ultra,
  title={Ultra-compact 8$\times$ 8 strictly-non-blocking Si-wire {PILOSS} switch},
  author={Suzuki, Keijiro and Tanizawa, Ken and Matsukawa, Takashi and Cong, Guangwei and Kim, Sang-Hun and Suda, Satoshi and Ohno, Morifumi and Chiba, Tadashi and Tadokoro, Hirofumi and Yanagihara, Masashi and others},
  journal={Optics Express},
  volume={22},
  number={4},
  pages={3887--3894},
  year={2014},
  doi = {10.1364/OE.22.003887},
  publisher={Optical Society of America}
}

@article{suzuki2019low,
  title={Low-insertion-loss and power-efficient 32$\times$ 32 silicon photonics switch with extremely high-$\Delta$ silica {PLC} connector},
  author={Suzuki, Keijiro and Konoike, Ryotaro and Hasegawa, Junichi and Suda, Satoshi and Matsuura, Hiroyuki and Ikeda, Kazuhiro and Namiki, Shu and Kawashima, Hitoshi},
  journal={Journal of Lightwave Technology},
  volume={37},
  number={1},
  pages={116--122},
  year={2019},
  doi = {10.1109/JLT.2018.2867575},
  publisher={OSA}
}

@article{maring2024versatile,
  title={A versatile single-photon-based quantum computing platform},
  author={Maring, Nicolas and Fyrillas, Andreas and Pont, Mathias and Ivanov, Edouard and Stepanov, Petr and Margaria, Nico and Hease, William and Pishchagin, Anton and Lema{\^\i}tre, Aristide and Sagnes, Isabelle and others},
  journal={Nature Photonics},
  volume={18},
  number={6},
  pages={603--609},
  year={2024},
  doi = {10.1038/s41566-024-01403-4},
  publisher={Nature Publishing Group UK London}
}

@article{motes2014scalable,
  title={Scalable boson sampling with time-bin encoding using a loop-based architecture},
  author={Motes, Keith R and Gilchrist, Alexei and Dowling, Jonathan P and Rohde, Peter P},
  journal={Physical Review Letters},
  volume={113},
  number={12},
  pages={120501},
  year={2014},
  doi = {10.1103/PhysRevLett.113.120501},
  publisher={APS}
}

@article{he2017time,
  title={Time-bin-encoded boson sampling with a single-photon device},
  author={He, Yu and Ding, X and Su, Z-E and Huang, H-L and Qin, J and Wang, C and Unsleber, S and Chen, C and Wang, H and He, Y-M and others},
  journal={Physical Review Letters},
  volume={118},
  number={19},
  pages={190501},
  year={2017},
  doi = {10.1103/PhysRevLett.118.190501},
  publisher={APS}
}

@article{fischer2021autonomous,
  title={Autonomous on-chip interferometry for reconfigurable optical waveform generation},
  author={Fischer, Bennet and Chemnitz, Mario and MacLellan, Benjamin and Roztocki, Piotr and Helsten, Robin and Wetzel, Benjamin and Little, Brent E and Chu, Sai T and Moss, David J and Aza{\~n}a, Jos{\'e} and others},
  journal={Optica},
  volume={8},
  number={10},
  pages={1268--1276},
  year={2021},
  doi = {10.1364/OPTICA.435435},
  publisher={Optica Publishing Group}
}

@article{bouchard2024programmable,
  title={Programmable photonic quantum circuits with ultrafast time-bin encoding},
  author={Bouchard, Fr{\'e}d{\'e}ric and Fenwick, Kate and Bonsma-Fisher, Kent and England, Duncan and Bustard, Philip J and Heshami, Khabat and Sussman, Benjamin},
  journal={Physical Review Letters},
  volume={133},
  number={9},
  pages={090601},
  year={2024},
  doi = {10.1103/PhysRevLett.133.090601},
  publisher={APS}
}

@article{monika2025quantum,
  title={Quantum state processing through controllable synthetic temporal photonic lattices},
  author={Monika, Monika and Nosrati, Farzam and George, Agnes and Sciara, Stefania and Fazili, Riza and Marques Muniz, Andr{\'e} Luiz and Bisianov, Arstan and Lo Franco, Rosario and Munro, William J and Chemnitz, Mario and others},
  journal={Nature Photonics},
  volume={19},
  number={1},
  pages={95--100},
  year={2025},
  doi = {10.1038/s41566-024-01546-4},
  publisher={Nature Publishing Group UK London}
}

@article{su2019conversion,
  title={Conversion of Gaussian states to non-Gaussian states using photon-number-resolving detectors},
  author={Su, Daiqin and Myers, Casey R and Sabapathy, Krishna Kumar},
  journal={Physical Review A},
  volume={100},
  number={5},
  pages={052301},
  year={2019},
  doi = {10.1103/PhysRevA.100.052301},
  publisher={APS}
}

@article{konno2024logical,
  title={Logical states for fault-tolerant quantum computation with propagating light},
  author={Konno, Shunya and Asavanant, Warit and Hanamura, Fumiya and Nagayoshi, Hironari and Fukui, Kosuke and Sakaguchi, Atsushi and Ide, Ryuhoh and China, Fumihiro and Yabuno, Masahiro and Miki, Shigehito and others},
  journal={Science},
  volume={383},
  number={6680},
  pages={289--293},
  year={2024},
  doi = {10.1126/science.adk7560},
  publisher={American Association for the Advancement of Science}
}

@article{aghaee2025scaling,
  title={Scaling and networking a modular photonic quantum computer},
  author={Aghaee Rad, H and Ainsworth, T and Alexander, RN and Altieri, B and Askarani, MF and Baby, R and Banchi, L and Baragiola, BQ and Bourassa, JE and Chadwick, RS and others},
  journal={Nature},
  pages={1--8},
  year={2025},
  doi = {10.1038/s41586-024-08406-9},
  publisher={Nature Publishing Group UK London}
}

@article{guha2025quantum,
  title={Quantum-enhanced quickest change detection of transmission loss},
  author={Guha, Saikat and John, Tiju Cherian and Gong, Zihao and Basu, Prithwish},
  journal={Physical Review Letters},
  volume={135},
  number={21},
  pages={210801},
  year={2025},
  doi={10.1103/czg5-3y3d},
  publisher={APS}
}

@article{oh2024classical,
  title={Classical algorithm for simulating experimental Gaussian boson sampling},
  author={Oh, Changhun and Liu, Minzhao and Alexeev, Yuri and Fefferman, Bill and Jiang, Liang},
  journal={Nature Physics},
  volume={20},
  number={9},
  pages={1461--1468},
  year={2024},
  doi = {10.1038/s41567-024-02535-8},
  publisher={Nature Publishing Group UK London}
}

@article{yu2023universal,
  title={A universal programmable Gaussian boson sampler for drug discovery},
  author={Yu, Shang and Zhong, Zhi-Peng and Fang, Yuhua and Patel, Raj B and Li, Qing-Peng and Liu, Wei and Li, Zhenghao and Xu, Liang and Sagona-Stophel, Steven and Mer, Ewan and others},
  journal={Nature Computational Science},
  volume={3},
  number={10},
  pages={839--848},
  year={2023},
  doi = {10.1038/s43588-023-00526-y},
  publisher={Nature Publishing Group US New York}
}

@article{pegoraro2024demonstration,
  title={Demonstration of a Photonic Time-Multiplexed C-NOT Gate},
  author={Pegoraro, Federico and Held, Philip and Lammers, Jonas and Brecht, Benjamin and Silberhorn, Christine},
  journal={arXiv preprint},
  doi = {arXiv:2412.02478},
  year={2024}
}

@article{sempere2022experimentally,
  title={Experimentally finding dense subgraphs using a time-bin encoded Gaussian boson sampling device},
  author={Sempere-Llagostera, S and Patel, RB and Walmsley, IA and Kolthammer, WS},
  journal={Physical Review X},
  volume={12},
  number={3},
  pages={031045},
  year={2022},
  doi = {10.1103/PhysRevX.12.031045},
  publisher={APS}
}

@article{russell2017direct,
  title={Direct dialling of Haar random unitary matrices},
  author={Russell, Nicholas J and Chakhmakhchyan, Levon and O’Brien, Jeremy L and Laing, Anthony},
  journal={New journal of physics},
  volume={19},
  number={3},
  pages={033007},
  year={2017},
  publisher={IOP Publishing}
}

@misc{Basani2024_CasOptAx,
  author = {Basani, J.},
  title = {Cascaded Optical Systems Approach to Neural Networks (CasOptAx)},
  year = {2024},
  publisher = {GitHub},
  journal = {GitHub repository},
  howpublished = {\url{https://github.com/JasvithBasani/CasOptAx}}
}

\end{document}